\documentclass[prb,aps, twocolumn, nofootinbib, superscriptaddress, longbibliography,  floatfix]{revtex4-2}

\pdfoutput=1
\usepackage[utf8]{inputenc}
\DeclareUnicodeCharacter{00A0}{ }
\usepackage{amsmath}
\usepackage{amssymb}
\usepackage{bbold}
\usepackage{mathtools}
\usepackage{physics}
\usepackage[makeroom]{cancel}
\usepackage{stmaryrd}

\usepackage{graphicx}
\usepackage{color}
\usepackage[english]{babel}
\usepackage{stmaryrd}
\usepackage{microtype}
\usepackage{float}

\usepackage[percent]{overpic}
\usepackage{tikz}
\usetikzlibrary{calc,shadows,matrix,positioning,patterns,decorations.pathreplacing}

\usepackage{hyperref}
\hypersetup{
	pdftoolbar=true,
	pdfmenubar=true,
	pdffitwindow=false,
	pdfstartview={FitH},
	pdftitle={},
	pdfauthor={Bauernfeind, Cao, Stoudenmire, Parcollet},
	pdfsubject={},
	pdfcreator={},
	pdfproducer={},
	pdfnewwindow=true,
	colorlinks=true,
	linkcolor=black,
	citecolor=blue,
	filecolor=magenta,
	urlcolor=blue
}

\usepackage{xr}

\newcommand{\avg}[1]{\ensuremath{\langle #1 \rangle}}
\newcommand{\ZZ}{\ensuremath{\mathcal{Z}}}

\usepackage{xcolor}

\definecolor{eggplant}{RGB}{148,33,147}
\definecolor{royalblue}{RGB}{4,51,255}
\definecolor{orange}{RGB}{255,147,0}
\definecolor{fern}{RGB}{79,143,0}

\newcommand{\CCQ}{Center for Computational Quantum Physics, Flatiron Institute, 162 5th Avenue, New York, NY 10010, USA}

\usepackage[addedmarkup=colored]{changes}

\begin{document}
\title{Minimally Entangled Typical Thermal States Algorithms for Finite Temperature Matsubara Green Functions.}

\author{Daniel Bauernfeind}
\affiliation{\CCQ}

\author{Xiaodong Cao}
\email{xcao@flatironinstitute.org}
\affiliation{\CCQ}

\author{E.\ Miles Stoudenmire}
\affiliation{\CCQ}

\author{Olivier Parcollet}
\affiliation{\CCQ}
\affiliation{Universit\'e Paris-Saclay, CNRS, CEA, Institut de Physique Th\'eorique, 91191, Gif-sur-Yvette, France}

\date{\today}

\begin{abstract}
We extend finite-temperature tensor network methods to compute Matsubara imaginary-time correlation functions,
building on the minimally entangled typical thermal states (METTS) and purification algorithms. 
While imaginary-time correlation functions are straightforward to formulate with these methods, 
care is needed to avoid convergence issues that would result from naive estimators.
As a benchmark, we study the single-band Anderson impurity model, 
even though the algorithm is quite general and applies to lattice models.
The special structure of the impurity model benchmark system and our choice of basis enable 
techniques such as reuse of high-probability METTS for increasing algorithm efficiency. 
The results are competitive with state-of-the-art continuous time Monte Carlo.
We discuss the behavior of computation time and error as a function of the number of purified sites in the Hamiltonian.

\end{abstract}

\maketitle

\section{Introduction}

The development of high precision quantum  many-body physics is a rapidly growing field, 
whose purpose is to obtain controlled solutions of strongly interacting quantum systems.
One direction of research are tensor network methods, which represent ground state many-body wavefunctions through a low-rank tensor
decomposition. The first and most famous approach of this class is the  
density matrix renormalization group (DMRG)~\cite{White_DMRG, UliRevModPhys.77.259, SCHOLLWOCK201196} which is very efficient for one dimensional systems and whose modern form is now understood in terms of matrix-product state (MPS) tensor networks.

The computation of real time dynamics and correlators in DMRG and tensor methods has been 
studied extensively, for a review see e.g. Ref.~\onlinecite{Paeckel}.
However, in order to study the thermodynamics and much of the equilibrium properties of a system, 
the complete and detailed knowledge of real time dynamics is not necessary. 
The Matsubara \emph{imaginary time} formalism in thermal equilibrium is sufficient and is
almost always easier computationally since imaginary time evolution involves
decaying exponentials. 
Indeed, many physical properties of a system can be derived from Matsubara self-energy, such as the Fermi liquid effective mass, quasiparticle lifetime, or the location of the Fermi surface.
Real frequency correlators can also be obtained to some extent at low energy 
from imaginary time results by using analytical continuation techniques, for example Padé approximants.

Quantum Monte Carlo (QMC) methods typically work in Matsubara imaginary time (with some notable recent exceptions 
\cite{InchwormCohenGullMillisRealTime2015,ProfumoRealTimeQMC, Gull_RealTimeMC, BertrandPRX, QQMCPRL}).
However they face limitations within some parameter ranges, especially at low temperatures. 
The most famous is the sign problem, which affects most QMC methods and manifests itself as an exponential 
increase of error bars at low temperatures.
Certain other QMC methods such as diagrammatic QMC \cite{ProkofievSvistunovDiagQMCPhysRevLett.81.2514}, which computes perturbative series with stochastic techniques, 
face difficulties with convergence as the interaction strength is increased, see e.g. Ref.~\onlinecite{BertrandPRX}.
It is therefore very interesting to explore the computation of Matsubara Green functions at finite temperature 
using tensor network methods for various systems, since these methods are not limited in the same way.

Apart from the imaginary time evolution itself, the main challenge in obtaining the Matsubara Green function 
is computing {\it finite temperature} versus ground state properties.  Within tensor network methods,
finite temperature systems can be accessed in two ways: {\it (i}) the  minimally entangled typical thermal states
(METTS) algorithm
\cite{White_Metts,Miles_Metts,Barthel_SymmetricMetts,Jing_HybridPurificationMetts,chung_2019minimally}
uses a sampling technique over tensor network states to produce the thermal
average; {\it (ii}) the purification technique
\cite{FeiguinPurification,Zwolak,Verstraete_purification,HauschildPurification}
rewrites the thermal trace as a average in a pure state in a larger space,
obtained by imaginary time evolution.

%While finite temperature correlation functions were computed previously, again, most of the focus was on real-time evolution via the purification~\cite{Purification_RT_Barthel,Purification_RT_Garst,Purification_RT_Meissner, Purification_RT_Karasch1, Purification_RT_Karasch2, Purification_RT_Karasch3,Purification_RT_Karasch4, Purification_RT_Karasch5, Purification_RT_Karasch6,Purification_RT_Karasch7, Purification_RT_Karasch8} or METTS~\cite{Metts_RT_Barthel,Metts_RT_Laeuchli, Metts_RT_Weichselbaum} and imaginary time correlators have not gotten a lot of attention.

In this paper, we show how to combine the METTS and the purification idea to compute the finite temperature Matsubara Green function with these tensor network approaches at any temperature.
As a first benchmark application, we use the single-site one-band Anderson quantum impurity model.
Indeed, quantum impurity models are a very interesting application for this algorithm in the context of dynamical mean field theory (DMFT) ~\cite{Georges_DMFT}.
DMFT and its extensions constitute a systematic approach to strongly correlated systems, not only for models but also in realistic computation \cite{KotliarRMP2006}.
The core of the technique consists in solving quantum impurity models, in self-consistently determined baths, at low temperatures.
These models are still complex many-body problems, and require advanced algorithms to solve them, such as continuous time QMC \cite{RubtsovCTQMC2005, cthyb2006, Gull_2008, CTQMC_RMP}  
the numerical renormalization group (NRG)~\cite{NRG_bulla, NRG_original},
DMRG \cite{JeckelmannPhysRevB.66.045114, PhysRevLett.93.246403,
   PhysRevB.72.113110,PhysRevB.72.113110,PhysRevB.52.R719,Schollwoeck_ImagTimeSolver,
   WolfChebyPRB2015, Schollwoeck_ImagTimeSolver_2,
   WolfClusterPhysRevB.90.115124, Zingl_ImagTimeSolver}, and recently more
general tensor network approaches~\cite{FTPS}.
Most MPS-based DMFT solvers have focused on real time evolution, with some notable
exceptions~\cite{Schollwoeck_ImagTimeSolver,Schollwoeck_ImagTimeSolver_2,Zingl_ImagTimeSolver}
but still at zero temperature.
The METTS/purification hybrid algorithm presented in this paper works at finite temperature.

This paper is structured as follows. In Sec.~\ref{sec:METTS} we briefly
introduce the original METTS algorithm as well as how to combine METTS and
purification and define the estimator to sample imaginary time correlators. In
Sec.~\ref{sec:ModelMethods} we benchmark the algorithm on the Anderson Impurity model
and discuss the balance between the METTS and purification in details.

% ========================================================================

\section{METTS for Imaginary time Correlation functions}\label{sec:METTS}

\subsection{The METTS algorithm}

The METTS algorithm~\cite{White_Metts, Miles_Metts} is a particular form of quantum Monte Carlo. 
It computes finite temperature thermal averages
\begin{align}\label{eq:thermalAvDef}
    \avg{A}= \frac{1}{\ZZ} \text{Tr}[e^{-\beta H} A]
\end{align} 
where $A$ is an observable, $\beta$ the inverse temperature,  $H$ the Hamiltonian and $\ZZ$ the partition function.
It is based on the observation that (\ref{eq:thermalAvDef}) can be rewritten as:
\begin{align}\label{eq:MettsBase}
\langle A \rangle &= \frac{1}{\ZZ} \Tr[ e^{-\beta H} A ]  \nonumber \\
			 &= \frac{1}{\ZZ}  \sum_i  \langle i | e^{-\beta H/2} A e^{-\beta H/2}  | i \rangle \nonumber \\
			 &= \frac{1}{\ZZ}  \sum_i P(i) \langle \phi_i | A | \phi_i \rangle 
\end{align}
where $\phi$ and $P$ are defined by 
\begin{align}\label{eq:defPhiandP}
| \phi_i \rangle &\equiv  e^{-\beta H/2} | i \rangle  / \sqrt{P(i)} \nonumber \\
P(i) &\equiv   \langle i | e^{-\beta H} | i \rangle \ \ .
\end{align}
The states $|\phi_i\rangle$ are non-orthogonal but normalized. When the
$|i\rangle$ are chosen to be a basis of classical, or unentangled product
states, the resulting $|\phi_i\rangle$ are called METTS. 
The METTS algorithm performs
the full imaginary time evolution $e^{-\beta H/2} | i \rangle $ at each step
and samples the basis states $\ket{i}$ with a Monte Carlo procedure. 
The algorithm starts with a state $\ket{i}$, time-evolves it in imaginary time up to
$\beta/2$, then computes observables as expectation values of the
$\ket{\phi_i}$ states. After that, a new basis state $\ket{j}$ is chosen with
probability $p(i\rightarrow j) =|\langle j | \phi_i \rangle |^2$. The collapse
into the new basis state $\ket{j}$ can be computed efficiently  \cite{Miles_Metts} by recursively
collapsing one site at a time.%
\footnote{Repeated use of the
   product rule of probabilities allows to decompose $ | \langle j | \phi_i
   \rangle |^2 \equiv P(j_1, j_2 \cdots j_N | \phi_i) = P(j_1| \phi_i) P(j_2 |
   j_1, \phi_i) P(j_3 | j_1, j_2, \phi_i) \cdots P(j_N | j_1 \cdots j_{N-1},
   \phi_i)$ proving that the local collapse is correct.}
Additionally, it is
worth noting that within the METTS algorithm, the newly sampled state $\ket{j}$
is always accepted as detailed balance is satisfied \cite{Miles_Metts}.

The METTS states $\ket{\phi_i}$ are non-trivial entangled states,
hence tensor networks, such as matrix product states, are used to
represent these states during and after the time evolution process. Note
however that the actual METTS algorithm presented above is independent of the
choice of wavefunction representation and could be implemented using any
representation that allows $\ket{\phi_i}$ to be computed efficiently.

\subsection{Imaginary time correlation functions}

The finite-temperature correlation function of two operators $A$ and $B$ in imaginary time for a system with Hamiltonian $H$ is defined as
(note the unusual sign convention for $G_{AB}$ used in this paper):
\begin{align}
    G_{AB}(\tau) &= \langle T_\tau A(\tau) B \rangle \nonumber \\
    A(\tau) &= e^{\tau H } A e^{-\tau H} \nonumber \\
    T_\tau A(\tau) B &= \begin{cases}
        +A(\tau) B \text{~~~~} \tau>0\\
        \pm B A(\tau)  \text{~~~~} \tau<0,
     \end{cases}
    \label{eq:Gle}
\end{align}
where the $\pm$ sign depends on whether the operators $A$ and $B$ are fermionic ($-$) or bosonic ($+$). Here we will be treating the fermionic case.
Because of the cyclic invariance of the thermal trace, $G_{AB}(\tau)$ can be written in two ways:
\begin{align}
%\begin{rcases}
    G_{AB}(\tau) &= \langle A(\tau) B \rangle  = \frac{1}{\ZZ} \sum_{i} P(i) \underbrace{\langle \phi_i |A(\tau) B| \phi_i \rangle}_{G_i^{+}(\tau)} \nonumber \\
    G_{AB}(\tau) &= \langle B A(\tau - \beta) \rangle  = \frac{1}{\ZZ} \sum_{i} P(i) \underbrace{\langle \phi_i |B A(\tau - \beta)| \phi_i \rangle}_{G_i^{-}(\tau)}\label{eq:GgrGle_mat}
    %\end{rcases}  \text{~~for $\tau \in [0,\beta]$.} 
\end{align}
where we have also written out each of the thermal averages as averages over the METTS labeled by $i$.
While the exact thermal expectation value of these two ways of writing
$G_{AB}(\tau)$ is necessarily the same, the estimators $G^+_i(\tau)$ and
$G^-_i(\tau)$ generally differ for each individual METTS.
Note that this is true even if $A$ and $B$ respects a symmetry, 
because the METTS states $\ket{\phi_i}$  can break this symmetry. 
Using the explicit form of $\ket{\phi_i}$ from (\ref{eq:defPhiandP}), we have
\begin{align}
    G_i^{+}(\tau) &= P(i)  \langle i | e^{-(\beta/2 - \tau) H}  A e^{ -\tau H } B e^{-\beta H/2}| i \rangle \nonumber \\ 
    G_i^{-}(\tau) &= P(i)  \langle i | e^{-\beta H/2} B e^{-\Tilde{\tau}H} A e^{-(\beta/2 -\Tilde{\tau})H} | i \rangle \nonumber\\
   \label{eq:mettsgf_full}
\end{align}
where $\Tilde{\tau} = \beta - \tau$.
For $\tau, \Tilde{\tau}>\beta/2$ these expressions can become very large because of the positive exponentials.
Therefore, we use the following {\it symmetric estimator} defined by:
\begin{equation}
    G_i(\tau) \equiv \left\{
                \begin{array}{ll}
                  G_i^+(\tau) \text{~~~ for $\tau<\beta/2$ } \\
                  G_i^-(\tau) \text{~~~ for $\tau>\beta/2$ ($\Tilde{\tau} < \beta/2$) } 
                \end{array}
              \right. \label{eq:symmetric_estimator}
\end{equation}
Moreover this symmetric estimator also allows to reuse partially
time-evolved states already computed during the calculation of the METTS
$\ket{\phi_i}$: if the METTS states and their normalization are kept during the computation, 
the time evolved states 
\begin{align}
    e^{-(\beta/2 - \tau) H} \ket{i},
\end{align}
for $\tau<\beta/2$ require no extra computation. 

The main drawback of the symmetric estimator Eq.~(\ref{eq:symmetric_estimator}) 
is to introduce a discontinuity at $\beta/2$ for any finite number of samples, 
since $G_i^+(\beta/2)$ and $G_i^-(\beta/2)$ are in general not equal.
We will discuss that point further in
Sec.~\ref{ssec:symmetricEstimator} on a concrete example.

%%%%%%%%%%%%%

\subsection{Hybrid METTS-Purification algorithm}

The METTS algorithm presented above has two main drawbacks, which we will now address by mixing it with the purification technique \cite{chung_2019minimally,Jing_HybridPurificationMetts}. First we will introduce purification then review how a controlled algorithm can be formulated which is hybrid of purification and METTS.

\subsubsection{Purification}

Let us start with a simple reminder of the purification idea.
For each physical (spinless) fermionic degree of freedom $P_i$ at site $i$ 
with two states $\ket{0}_{P_i}$ and $\ket{1}_{P_i}$, we take the tensor product with an auxiliary or ``ancilla'' degree of freedom
$A_i$ with two states $\ket{0}_{A_i}$ and $\ket{1}_{A_i}$ and define the maximally entangled state $S_i$ as 
\begin{equation}\label{eq:purificationMaxEntStateExp}
\ket{S_i} \equiv \frac{1}{\sqrt{2}} \left( \ket{0}_{P_i} \otimes \ket{1}_{A_i} + \ket{1}_{P_i} \otimes \ket{0}_{A_i} \right)
\end{equation}
such that 
\begin{equation} \label{eq:purificationMaxEntState}
\Tr_{A_i} \ket{S_i} \bra{S_i} = \mathbb{1}_{P_i} 
\end{equation}
where $P_i$ or $A_i$ refer to physical and auxiliary degrees of freedom on site $i$ respectively
($\Tr_{A_i}$ is the trace over the auxiliary degrees of freedom).

The density matrix can then be rewritten as 
\begin{align}
    \rho &= \frac{1}{\ZZ} e^{-\beta H/2} \cdot \mathbb{1}_{P} \cdot e^{-\beta H/2} \nonumber \\
    &= \frac{1}{\ZZ} e^{-\beta  H/2} \Tr_{A} \bigotimes_i \bigg(  \ket{S_i} \bra{S_i} \bigg) e^{-\beta H/2} \nonumber \\
    &= \frac{1}{\ZZ} \Tr_A \left[ \underbrace{\left(e^{-\beta  H/2}_P \otimes \mathbb{1}_A \right ) \ket{S}}_{\ket{\psi_\beta}} \bra{S} \left(e^{-\beta  H/2}_P \otimes \mathbb{1}_A \right ) \right ] \nonumber \\
    &=  \frac{1}{\ZZ} \Tr_A \ket{\psi_\beta} \bra{\psi_\beta}.
\end{align}
where $\ket{S} \equiv \otimes_i \ket{S_i}$,  $\Tr_A$ is the trace over
auxiliary degrees of freedom and the subscript $P$,(resp. $A$) means that the operator
acts on the physical (resp. auxiliary) space.  We also used $\mathbb{1}_P
\equiv \bigotimes_i \mathbb{1}_{P_i}$ and (\ref{eq:purificationMaxEntState}).
In other words, at the cost of doubling the dimension of the Hilbert space, 
the introduction of the ancilla states
lets us rewrite the density matrix as partial trace of a pure state density matrix, 
which can be obtained from a time evolution from $\ket{S}$ using pure-state time evolution methods. 

\subsubsection{Hybrid algorithm}

The purification technique has two main advantages over the original METTS algorithm.

First, although the METTS algorithm implemented with MPS is highly efficient at low temperature
and scales similarly to DMRG, its sampling becomes less efficient at high temperatures 
compared to the purification approach.

Secondly, because METTS updates remain in a fixed quantum-number sector corresponding 
to the symmetries of the Hamiltonian, one cannot straightforwardly use quantum numbers to simulate the
grand canonical ensemble efficiently~\cite{Miles_Metts, Barthel_SymmetricMetts}. 
On the contrary, using purification, the maximally entangled state $\ket{S}$ allows particle numbers to fluctuate
on the physical sites while keeping the total number of particles conserved in
the total space (physical and auxiliary). 

Remarkably, one can mix the two algorithms and purify only some fermionic sites, creating an hybrid METTS-purification algorithm \cite{Jing_HybridPurificationMetts, chung_2019minimally}.
The default METTS algorithm discussed above corresponds to purifying no sites.
When every site is purified, no METTS sampling is necessary as the purified state computes the density matrix of the whole system.
This hybrid algorithm is simple to implement from the original METTS, as the collapse step only needs to be modified slightly: 
purified sites are ignored at first (not collapsed) and after the collapse on all other sites
is completed, purified and ancilla pairs are reinitialized in the
maximally entangled state given by Eq.~(\ref{eq:purificationMaxEntStateExp})
\cite{Jing_HybridPurificationMetts}. 

The choice of the purified sites is a delicate question: it is a free parameter of the algorithm and
 is certainly model dependent. Two extreme limits are clear.
At high temperatures, purifying all of the sites is the most efficient choice. This can
be understood by observing that at infinite temperature a full purification immediately gives the 
exact thermal density matrix without any time evolution or sampling \cite{SCHOLLWOCK201196}. 
In contrast, at zero temperature the thermal density matrix is the tensor product
$\ket{\psi_0} \bra{\psi_0}$ where the ground state $\ket{\psi_0}$ can be represented with a single METTS and thus 
no purification is needed.
Away from these limits, purifying additional sites decreases the amount of sampling
that is necessary while increasing the cost of computing each sample. 
In quantum impurity models, a physically motivated choice is to purify only the sites of the 
impurity and the bath sites with the largest thermal fluctuations.
This approach will be investigated in detail in the next section.

%These issues can be circumvented by a hybrid METTS-purification scheme proposed in Refs.~\onlinecite,
%which is an unbiased and controlled method balancing the number of samples needed and the cost of calculating each sample. 

%When the total particle number $N$ and spin $S^z$ is conserved, 
%the local Hilbert space $\{ |0\rangle, |\uparrow\rangle, |\downarrow\rangle,
%|\uparrow\downarrow\rangle\}$ can be split into a product of two spinless ones
%$\{|0\rangle, |1\rangle \}_{\uparrow} \bigotimes \{|0\rangle, |1\rangle
%\}_{\downarrow}$. And for each spinless fermion degrees of freedom, we will
%collapse to the particle number basis while purifying some of the
%sites~\cite{chung_2019minimally,Jing_HybridPurificationMetts}. In practice, a set of purified sites is chosen for which the Hilbert space is doubled, while every other site is treated as in the  METTS algorithm.

%Increasing the fraction of sites which are purified decreases the size of the
%sampling space until eventually sampling is no longer necessary if every site
%is purified. In that limit, the mixed METTS-purification approach reduces to
%the regular purification algorithm which computes the full density matrix. This
%can be seen by noting that the infinite temperature density matrix can be
%viewed as an imaginary time evolution from the (unnormalized) infinite
%temperature state of the physical system $\mathbb{1} \equiv \bigotimes_i
%\mathbb{1}_{P_i}$ and we can rewrite it as:

%%%%%%%%%%%%%%%%%%%%%%%%%%%%%%%%%%%%%%%%%%%%
%%%%%%%% MODEL 
%%%%%%%%%%%%%%%%%%%%%%%%%%%%%%%%%%%%%%%%%%%%5

\section{Application to the Anderson impurity model}\label{sec:ModelMethods}

\subsection{Model} 

\begin{figure}[ht!]
    \centering
    \includegraphics[scale=0.6]{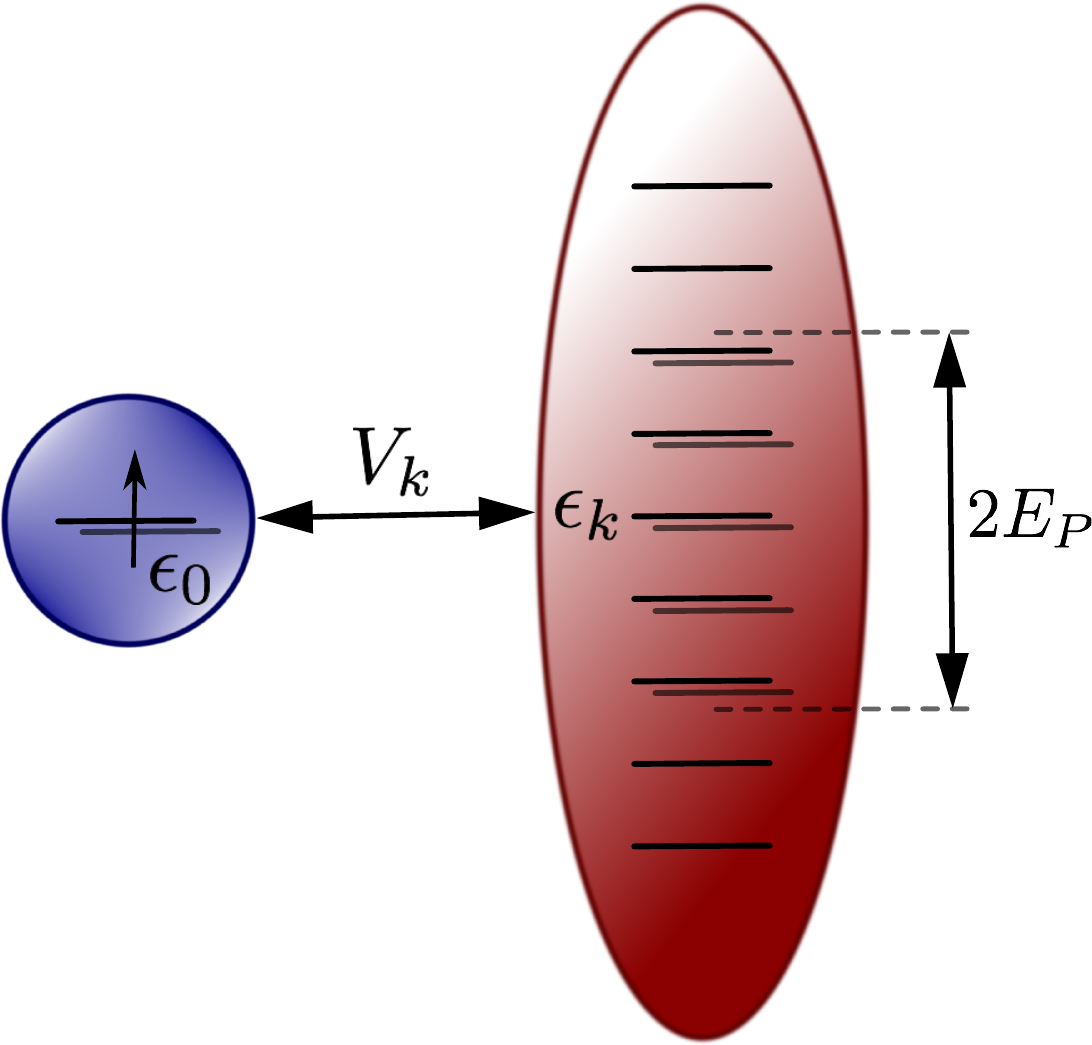}
    \caption{Schematic depiction of the impurity model used in this study. Purified sites are shown as double lines. We choose to always purify the impurity, as well as every bath site with an on-site energy $|\epsilon_k| < E_P$. }
    \label{fig:impmodels_schematic}
\end{figure}

We benchmark the hybrid METTS/purification algorithm on a single-band Anderson impurity model, defined by the Hamiltonian:
\begin{align}\label{eq:Himp}
        H = & \ U n_\uparrow n_\downarrow + \epsilon_0 \left( n_\uparrow + n_\downarrow \right) \nonumber \\ 
          + &\sum_{k\sigma} \epsilon_k n_{k\sigma} + V_k \left( d^\dagger_\sigma c_{k\sigma} +h.c. \right)
\end{align}
where $d^\dagger_\sigma$ and $d_\sigma$ are impurity creation and annihilation
operators of impurity electrons with spin $\sigma$. The $n_\sigma =
d^\dagger_\sigma d_\sigma$ are the corresponding particle number operators.
Similarly, $c^\dagger_{k\sigma}$ and $c_{k\sigma}$ create and annihilate a
spin-$\sigma$ electron at bath orbital $k$. 

In general, the effect of the bath onto the impurity is described by the hybridization function:
\begin{equation}
    \Delta(i\omega_n) = \sum_k \frac{V_k^2}{i\omega_n - \epsilon_k}\label{eq:delta_z}
\end{equation}
which, in many applications, is a general (causal) function that would require an infinite number of bath sites to represent.
Algorithms based on an Hamiltonian representation (like exact diagonalization, DMRG, NRG) require an approximation of $\Delta$ by a finite bath in Eq.~(\ref{eq:Himp}). 
The precise way to discretize the bath depends on the algorithm in question.
Here we use a bath with $N_b$ linearly spaced bath
energies $\epsilon_k$~\cite{FTPS, WolfStar, NRG_bulla} (see
Appendix~\ref{app:discretization}).
The energy resolution of our method is given by the difference between
consecutive bath energies $\epsilon_k$. 
In the following, we choose a
semi-circular bath of half bandwidth $D=2$ corresponding to a tight-binding
model on a infinite coordination Bethe lattice with hopping parameter $t=1$. 
We use $N_b = 109$ sites and we do {\it not} vary the bath size with $\beta$.

\subsection{Algorithm} 

We apply the algorithm presented in Section~\ref{sec:METTS}
to compute imaginary-time dependent properties of the Anderson impurity model defined above.
Before presenting the results, let us discuss aspects of the algorithm implementation which are
specific to this system.

First a physically-motivated criterion is needed to choose which sites to purify. Here we choose to purify sites 
likely to have the largest charge fluctuations, namely the impurity site and bath sites with energy $|\epsilon_k| < E_P$ for
some threshold energy $E_P$, as represented on Fig.~\ref{fig:impmodels_schematic}.
%Indeed, thermal fluctuations affect sites with small $|\epsilon_k|$. Other sites will be more likely empty or full.
It is also natural to purify the impurity itself, because of the Kondo spin flips at low energy.

In order to represent the METTS states, we use the fork tensor product states (FTPS) \cite{FTPS} tensor network, 
which allows the use of a large number of bath sites ($N_b \sim 100$). 
The imaginary time evolution is done with the time dependent
variational principle (TDVP) \cite{TDVP, TDVP_TTN}. While generally this
approach is very accurate even when using large time steps $\Delta \tau$, it
can produce incorrect results when bond dimension of the matrix product state is very
small as generically happens at the beginning of each step of the METTS algorithm. 
Therefore, when starting from a product state $\ket{i}$, we first perform
a small number of tDMRG~\cite{FTPS} (or Trotter-gate evolution) steps to increase the
tensor dimensions and then switch to TDVP. In this work, we use a few tDMRG steps of size $\Delta \tau_{tDMRG} =
0.05$. Recently, a basis extension for TDVP was proposed
which should allow the use of only TDVP for the time evolution \cite{Mingru_BasisExtensionTDVP}. 
We however leave this extension for future work. 

In practice, we observe that the METTS algorithm revisits the same product states quite often.
Depending on $E_P$ and the sample size, $10\% - 90\%$ of the
METTS appear at least twice. This behavior is quite specific to the quantum impurity problem we study here and the basis used to represent it, and would not be expected to occur in a generic model when working in a real-space basis.
Here we can take advantage of it by saving each METTS we have previously computed in memory, 
reusing them if the same initial product state is encountered, thus skipping redundant imaginary time evolution
calculations which are the most costly part of the algorithm. 

\subsection{Results}\label{sec:results}

We now present our results for the impurity single-particle Green function (omitting
the spin index)
\begin{align}
    G(\tau) = \langle T_\tau d(\tau) d^\dagger \rangle,
\end{align}
which are computed by first performing 20 steps of tDMRG with a time step of $\Delta\tau_{\text{tDMRG}}=0.05$, and then evolving to $\beta/2$ by TDVP with a time step of $\Delta\tau_{\text{TDVP}}=0.5$. We use a truncation error cutoff of $10^{-8}$ for both stages.

\begin{figure}[hb!]
    \centering
    \includegraphics{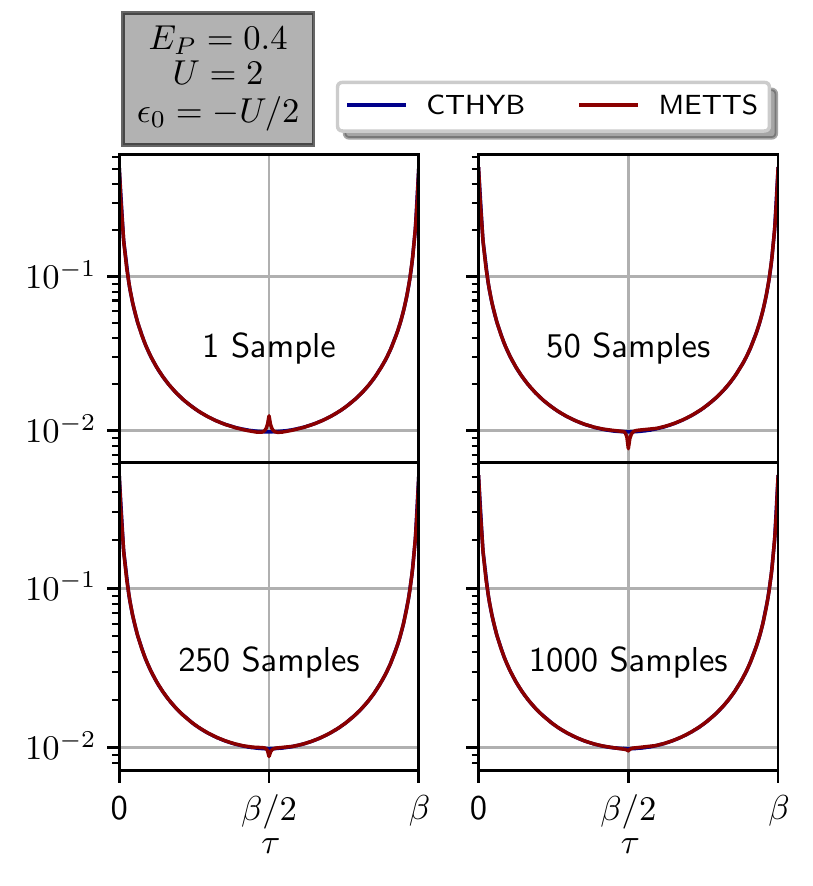}
    \includegraphics[]{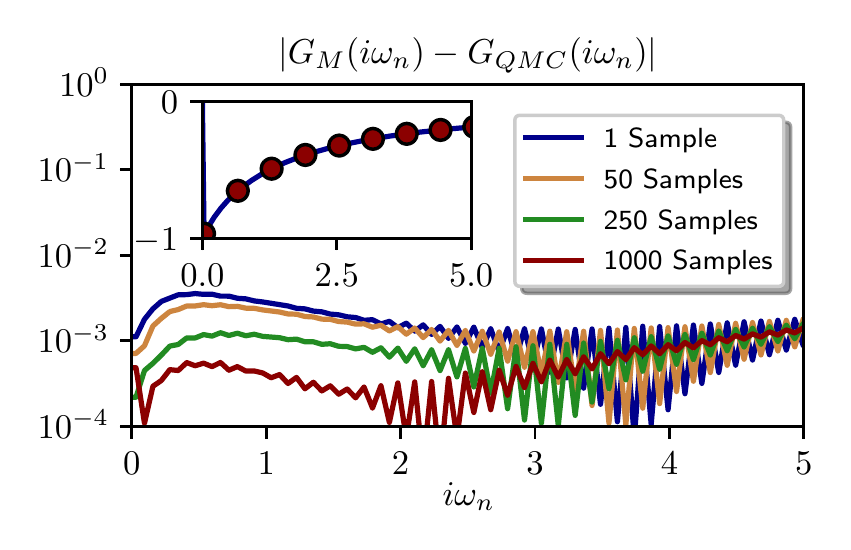}
    \caption{
       (Upper panel) Imaginary-time Green function $G(\tau)$ vs $\tau$ computed by the 
       METTS algorithm for different number of samples, compared to the QMC/CTHYB solution  
       for the same finite discretized bath.
       For the single-sample panel, we use the highest probability METTS.
       The fraction of METTS reused was
       between $0.6$ (50 samples, 20 unique METTS were computed) and $0.8$
       (1000 samples, 200 unique METTS were computed). 
       (Lower panel) Difference between the METTS ($G_M$) and the QMC result ($G_{QMC}$).
       (Inset) METTS (red, with 1000 samples) and QMC results (blue) at low frequencies.
       For all graphs, we used $U=2$, $\beta=100$, $E_P = 0.4$ and $\epsilon_0 = -U/2$ (particle-hole
       symmetry).
       }
    \label{fig:U2Convergence}
\end{figure}

\subsubsection{Benchmark and convergence}

In Fig.~\ref{fig:U2Convergence} we show the Green
function, for $U=2$, at low temperature $ \beta = 100$, 
in the particle-hole-symmetric case, 
with a purification window energy $E_P = 0.4$. 
We compare it to a QMC (CTHYB) solution for the exactly the same finite discretized bath, 
using the TRIQS/CTHYB code \cite{Seth_2016}.
The agreement is excellent even with a small sample size.
In Matsubara frequencies (Fig.~\ref{fig:U2Convergence}, lower panel), 
the agreement is also very good even at small frequencies, with an  error of order of $10^{-3}$.
Additional results for larger interactions and particle-hole asymmetric impurity models can be found in
Appendix~\ref{app:additionalresults}.

We observe that one of the METTS has a very high probability $P(i)/\ZZ$
and dominates the whole distribution:
 $P(i)/\ZZ \approx 70\%$ in the present case.
Using this state alone (Fig.~\ref{fig:U2Convergence}, first panel) already yields an excellent 
approximation of the Green function.
Unsurprisingly, this METTS corresponds to having a filled Fermi sea the set of non-purified bath sites.
At this low temperature, purifying about $20\%$ ($E_P = 0.4)$ of the bath sites captures most
thermal fluctuations already and there is no need to purify every site.

\begin{figure}[ht!]
    \centering
    \includegraphics{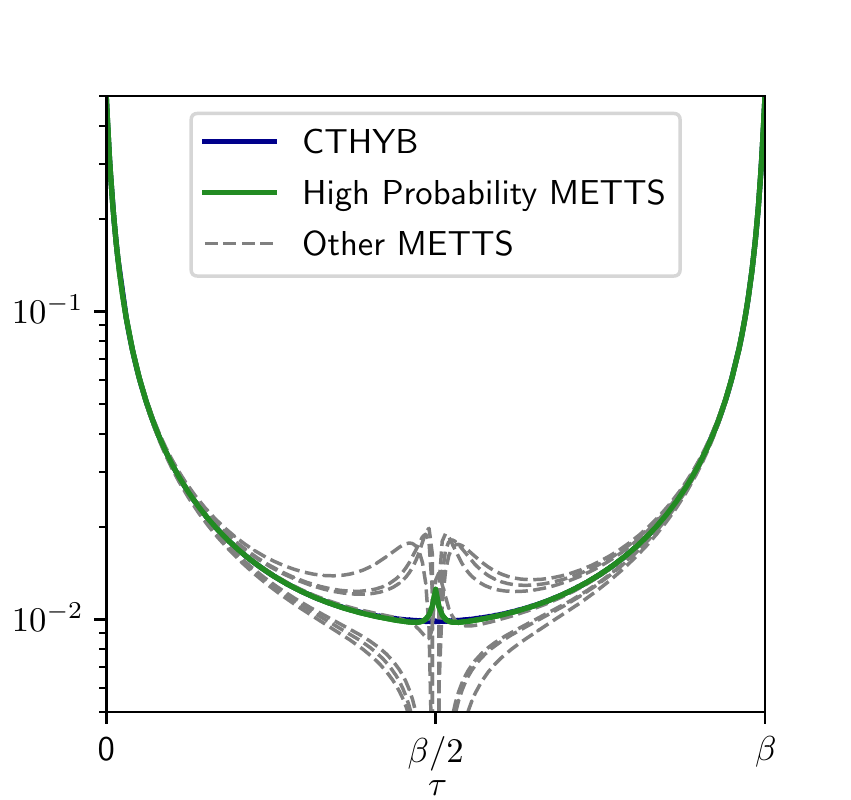}
    \caption{
       First 8 METTS Green functions in the METTS Markov chain shown in Fig.~\ref{fig:U2Convergence} compared to the CTHYB reference calculation. 
       The highest probability METTS ($P(i)/\ZZ \approx 0.7$) is plotted in green. 
       }
    \label{fig:U2_GFsample}
\end{figure}

The contribution of other METTS states to $G_i^\pm(\tau)$ are shown in
Fig.~\ref{fig:U2_GFsample}. They display a very large variance around $\beta/2$ (even with sign change).
This explains why in Fig.~\ref{fig:U2Convergence} the Green function does not appear to
improve when increasing the sample size from $1$ to $50$. 
In Figure~\ref{fig:histogram} in Appendix~\ref{app:histogram}, we show detailed histograms for $G(\beta/2)$ for various temperatures.

\begin{figure*}[ht!]
    \centering
    \includegraphics{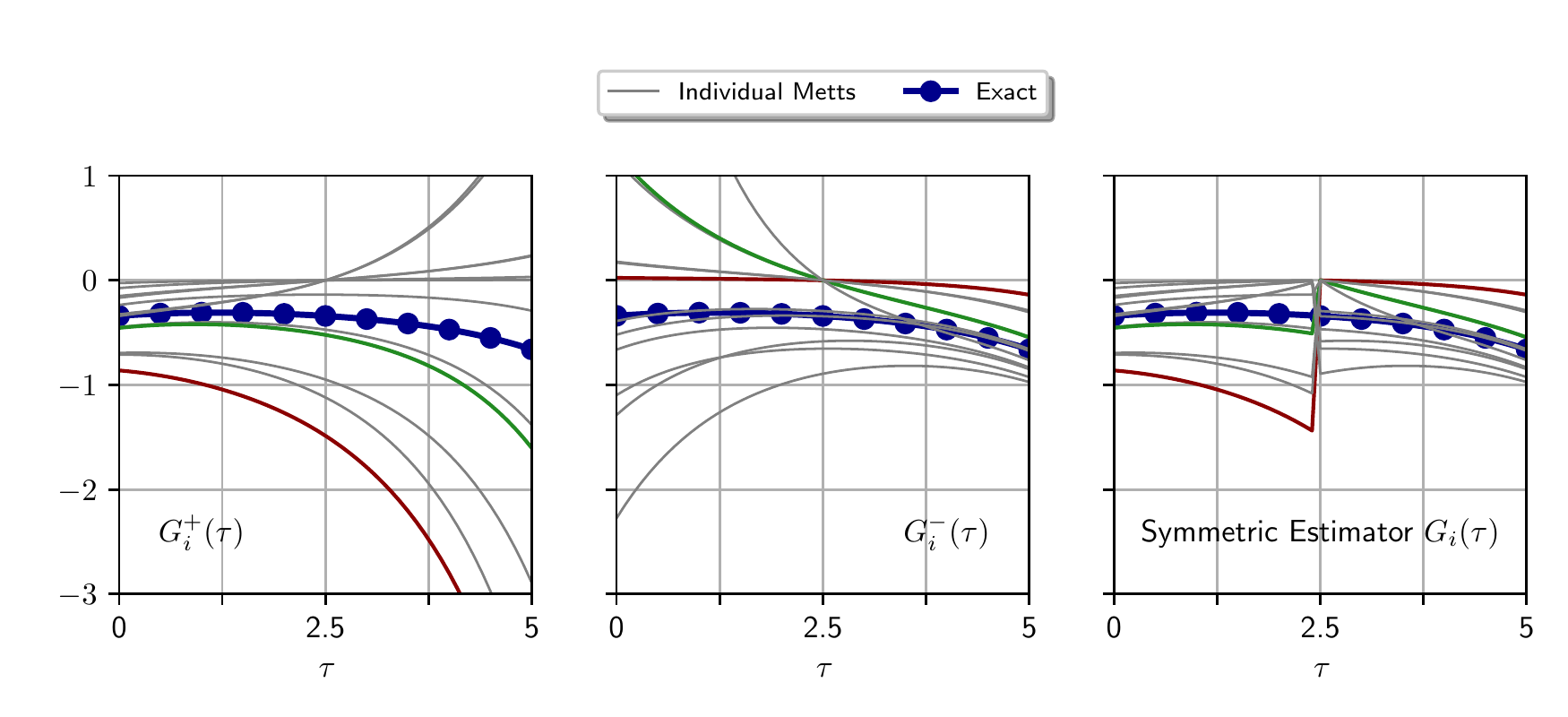}
    \caption{
       Typical $G_i^+(\tau)$, $G_i^+(\tau)$ and symmetric estimator $G_i(\tau)$ values
       for a small, non-interacting ($U=0$) impurity model with seven bath sites at $\beta=5$.
       No sites are purified.
       The plots show the $10$ most probable METTS obtained from particle number basis states $\ket{i}$. 
       The exact solution (sum of all METTS) is shown with symbols.
       Two specific METTS are highlighted in red and green, to show that for the same METTS either $G^+_i(\tau)$
       or $G^{-}_i(\tau)$ can deviate significantly from the average, highlighting the benefit of using the symmetric
       estimator.
       %The value for two METTS is represented in red and green, to compare their value for both $G_i^\pm$
       %$G_i^+(\tau)$ is generically stable (not exponentially growing) 
       %\miles{Is it really exponentially growing or just further from the average?} for $\tau<\beta/2$ and
       %gets large for $\tau > \beta/2$. Similarly $G_i^-(\tau)$ is
       %stable for $\tau>\beta/2$.
       } 
    \label{fig:smallsystem}
\end{figure*}
\subsubsection{Error around $\beta/2$}\label{ssec:symmetricEstimator}
The error in $G(\tau)$ is mostly concentrated in a relatively small time window around $\beta/2$.
It decreases with sampling, but slowly.
The reason for that is first, as mentioned above, $G_i^+(\beta/2)$ and $G_i^-(\beta/2)$ differ.
This yields a discontinuity in the symmetric estimator, which cancels only on average.
Second, the variance of $G(\tau)$ is larger around $\beta/2$ in the sampling.

In order to better understand this phenomenon, we solve a small non-interacting impurity model
exactly using a full diagonalization and we do an exact enumeration of all METTS $\ket{\phi}$, without approximation.

The results are presented in Fig. \ref{fig:smallsystem},
where we plot  $G_i^+ (\tau)$,
$G_i^- (\tau)$ and the symmetric estimator for the 10 METTS with the highest probability
$P(i)/\ZZ$. The METTS with the next-highest probability is $P(i)/Z = 0.007$.
Both $G_i^+(\tau)$ (resp. $G_i^-(\tau)$) become very large for at least some of the METTS
for  $\tau > \beta/2$ (resp. $\tau < \beta/2$).
As expected, the symmetric estimator Eq.~(\ref{eq:symmetric_estimator}) 
does not become large, but introduces a discontinuity in the contribution from each METTS at $\beta/2$. 
In both $G_i^\pm$, the sample variance is greatest around $\beta/2$, as in Fig.~\ref{fig:U2Convergence}.

%A small sample size of
%$50$ or even $1$ which previously gave very accurate results would fail
%drastically in this case. The reason for that is not only the small system
%size, but also the small value of $\beta$ as well as not purifying any site. In
%the next section we will thus discuss the interplay between between these two
%parameters $\beta$ and $E_P$ defining how many sites are purified.

%Even if it were possible to compute $G^+_i(\tau)$ or $G^-_i(\tau)$ for the
%whole $\tau$-axis, sampling from them would not result in a decrease of sample
%variance at $\beta/2$ (which was the main source of error in
%Fig.~\ref{fig:U2Convergence}) but would only \emph{increase} the variance in
%the $\tau$-region where these quantities diverge.

%%%%%%% Purification window CHOICE

\subsubsection{Choice of Purification Window}\label{ssec:EpVSbeta}

%\begin{figure}[ht!]
    %\centering
    %\includegraphics[scale = 0.8]{figures/TimeSD_vs_Ep.pdf}
    %\caption{\DB{This plot or the one above!?} }
    %\label{fig:VarTime_Sample_2}
%\end{figure}

Let us now examine the effect of the purification window energy $E_P$
as a function of $\beta$.
As discussed above, in this impurity model, a single METTS dominates with a high probability.
The contribution of excitations to the Green function are quite different as shown in
Fig.~\ref{fig:U2_GFsample}. They can even change sign around $\beta/2$.
Sampling from such a distribution is clearly hard
(see App.~\ref{app:histogram} for the shape of the distribution), which is an incentive to increase $E_P$.

\begin{figure}[ht!]
    \centering
    \includegraphics{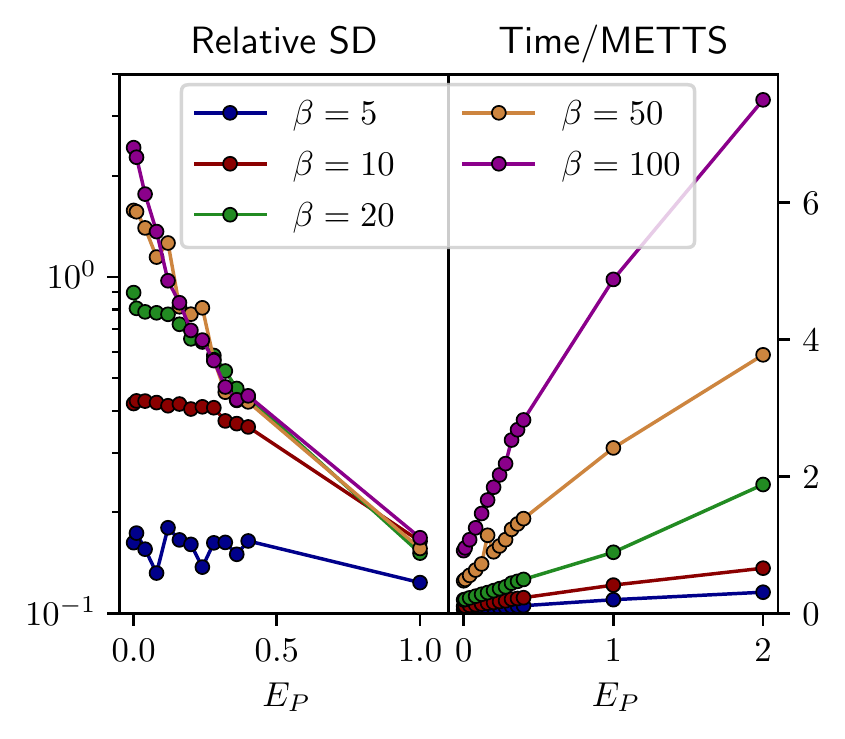}
    \caption{(Left) Relative standard deviation at $\beta/2$ as a function of purification energy scale $E_P$. (Right) Wall time in ($1000s$) to compute a single METTS also as a function of $E_P$. The first two points which are both very close to $E_P=0$ correspond to only purifying the impurity and purifying the impurity and the zero-energy bath site. Beyond that for $E_P \leq 0.4$, each data point has two additional bath sites purified (one site for positive and negative $\epsilon_k$). Finally note that $E_P=2$ means that every site is purified and no sampling is necessary at all. Therefore this point is excluded from the plot on the left. }
    \label{fig:VarTime_Sample}
\end{figure}

The optimal choice of the purification window energy $E_P$ is therefore a delicate balance.
When $E_P$ increases, the computation cost of a METTS increases, but the variance of the distribution decreases.
In order to quantify this balance more precisely, we show on  Fig.~\ref{fig:VarTime_Sample}
the relative standard deviation (RSD) of $G(\beta/2$) (left panel) and the computation time per METTS (right panel), 
for the $U=0$ case (note that non-interacting models are non-trivial for tensor network approaches).
Apart from small $\beta$, where it is nearly independent of $E_P$,
the RSD is similar for all $\beta$ (note however that for large $\beta$, $G(\beta/2)$ is very small, so the RSD is large
even though the results improve with increasing $\beta$).
The computing time for a single sample increases with $E_P$
even without taking into account that one would have to actually decrease the
truncated weight to get similar accuracies. The difference in computation time
between different choices of $E_P$ can be large. For example, a
factor of about three can be gained at all values of $\beta$ if the highest
probability METTS of $E_P=0.4$ is used instead of the full purification.

\begin{figure*}[ht!]
    \centering
    \includegraphics{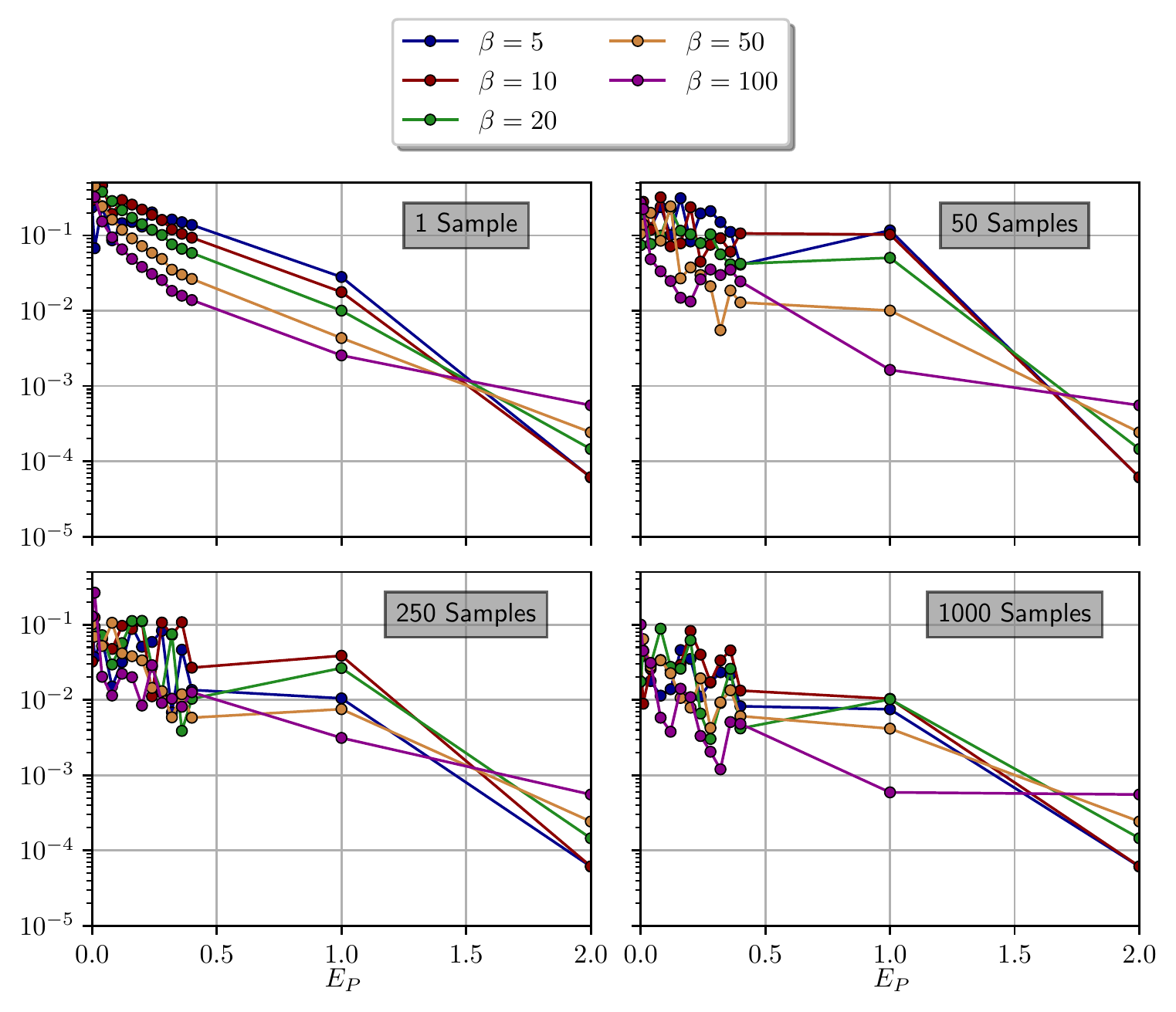}
    \caption{Error defined as $\sum_i |G_{Ex}(\tau_i)- G_{METTS}(\tau_i)|$ for a non-interacting impurity model. For the single sample, we again chose the highest probability METTS. Note that for $\beta=5$ and low $E_P$ the highest probability METTS is not well defined for the sample sizes studied. Again, the $E_P=2$ values correspond to the full purification method and for those data points, no sampling is needed.  }
    \label{fig:U0_errors}
\end{figure*}

Finally, we show in Fig.~\ref{fig:U0_errors} the error of the Green function $G(\tau)$ for various
sample sizes, values of $\beta$ and $E_P$. Starting from a single sample where
we always choose the highest probability METTS, the Green function converges to
the exact result by increasing the sample size. Nevertheless, in some 
cases a finite amount of sampling can even result in a worse Green
function compared to just using the highest-probability METTS as seen for $\beta=20$ and $E_P=1$, for example.

%%%%%%%%%%%%%%%%%%%%%%

\subsubsection{Reuse of METTS state}
\begin{figure}[ht!]
    \centering
    \includegraphics{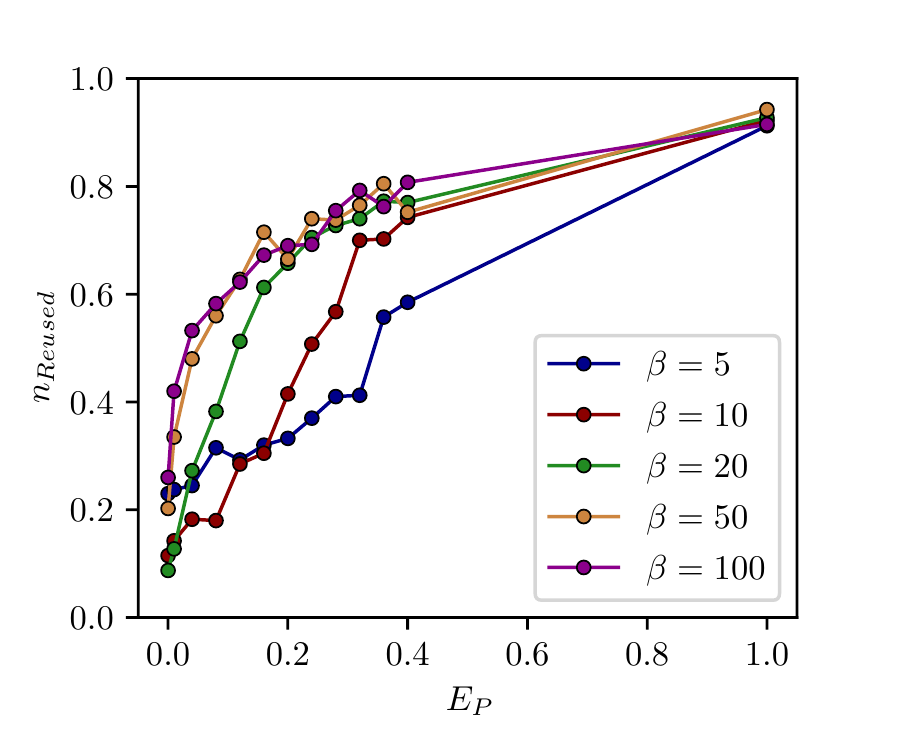}
    \caption{ Fraction of METTS reused for a sample of size $400$ as a function of $E_P$. At $E_P = 1$ half the bath sites are purified and a $\beta$-independent fraction of about $90\%$ of the METTS appear at least twice. From those, the highest probability METTS is the main contributor followed by the single- and two-particle excitations etc. }
    \label{fig:NReused}
\end{figure}

A crucial factor to reduce overall computation times for this class of system 
is to reuse the already computed METTS. 
In Fig.~\ref{fig:NReused} we plot the fraction of METTS that appeared at least twice and 
for which we were able to reuse the Green function and METTS $\ket{\phi_i}$. Like the RSD, this
fraction does not strongly depend on $\beta$ as soon as $E_P$ exceeds some
$\beta$-dependent value (for $\beta=10$, $E_P = 0.32$). By purifying half the
bath ($E_P = 1$), a $\beta$ independent fraction of about $90\%$ of the sample
can be reused. From these 90$\%$, the by far largest contribution comes of
course from the highest probability METTS with a small number $\mathcal{O}(1)$
of single-particle excitations that appear multiple times.

%%%%%%%%%%%%%%%%%%%%%%%%%%%   CONCLUSION %%%%%%%%%%%%%%%%%%%%%%%%%%%%%%%%%%%%

\section{Conclusion}

We have presented an extension of finite-temperature tensor network algorithms  
to obtain imaginary-time correlation functions, such as the Matsubara single-particle Green function $G(\tau)$.
Though the steps to obtain $G(\tau)$ are straightforward,
care is needed in devising estimators with good convergence properties.
Benchmarking the algorithm on the single-band Anderson quantum impurity model yields
excellent results when compared to well-established quantum Monte Carlo algorithms, 
with a relative accuracy of order $\sim 10^{-3}$ at low Matsubara frequencies.

For the Anderson model, we purified sites with the lowest magnitude single-site energies 
while sampling the rest as in the METTS algorithm. One improvement could be to solve the finite-temperature system in the non-interacting limit and purify sites with fractional occupations, rather than using a simple energy criterion.  
In more general future applications the choice of sites to purify will depend on the specific system. 
Also the balance between METTS sampling and purification depends on
the computing resources available: purifying fewer sites makes each METTS less costly to compute but requires
 more to be sampled. Yet this sampling overhead can parallelized away with more nodes available.

An interesting future direction is applying the methods presented here to real space systems,
for example lattice models of fermions such as the Hubbard model. 
METTS and purification already provide accurate results for the strongly-correlated
regime of the Hubbard model at low temperatures \cite{wietek2020stripes,wietek2021mott}.
Having access to the Matsubara Green function would yield additional insight into phenomena 
like the metal-insulator transition and Fermi surface properties.
In those settings, certain aspects particular to the benchmark impurity system we used here
would need to be adapted, especially the choice of which sites to purify.

Another valuable topic to investigate further would be improvements to the algorithm or estimator to 
enhance convergence near $\tau \approx \beta/2$. Though we still obtained accurate results near this point, it
is the slowest to converge due to a discontinuity in the contribution from each METTS as discussed in Section~\ref{sec:ModelMethods}. Finding an estimator or modified algorithm that would  eliminate such discontinuities is an open interesting question.

Finally, the effectiveness of METTS as an imaginary-time solver for more complex, multi-orbital impurity models
will be important to investigate. 
In particular, the optimum balance between purification and METTS could be significantly modified in the multi-orbital case.
Nevertheless, since tensor network methods are insensitive to the sign problem which limits usual continuous time quantum Monte Carlo, 
the algorithm we present here could be a method of choice to solve realistic DMFT equations at low temperatures, 
especially in the presence of strong spin-orbit coupling.

\appendix
\section{Bath Discretization}\label{app:discretization}
The descretization of the bath starts with a given bath spectral function $-\frac{1}{\pi} \Im \Delta(z = \omega +i0^+)$ (see Eq.~(\ref{eq:delta_z}). First, we find the energy range $[\omega_L, \omega_U]$ in which the bath spectral function is non-zero. We then split the energy axis into $N_b$ linearly spaced intervals of size $\Delta \epsilon = \frac{\omega_U - \omega_L}{N_b}$ such that the $k$-th interval $I_k = [ \omega_L + (k-1) \Delta \epsilon, \omega_L + k \Delta \epsilon ]$. We then place a bath site in the middle of this interval and compute the bath parameters via:
\begin{align}
    V_k^2 = -\frac{1}{\pi} \int_{I_k} d\omega \Im \Delta(\omega) \nonumber \\
    \epsilon_k = \omega_L + (k-0.5) \Delta \epsilon.
\end{align}

\section{Tensor Network}
In this appendix we give some additional details about the tensor network itself and the methods used to perform the time evolutions.

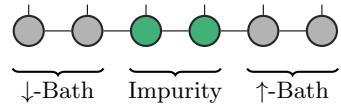
\begin{figure}[ht!]
   \centering
   \begin{tikzpicture}
    
   \definecolor{DanBlue}{RGB}{66,133,244}
   \definecolor{DanRed}{RGB}{219,68,55}
   \definecolor{DanGreen}{RGB}{15,157,88}
   \definecolor{DanYellow}{RGB}{244,160,0}

    \tikzstyle{old}=[	circle, thick, minimum size=0.4cm,
    				draw=black!85, fill=DanBlue!80, inner sep=0pt,  text width=4mm]

     \tikzstyle{new}=[	circle, thick, minimum size=0.4cm,
    				draw=black!85, fill=DanRed!80, inner sep=0pt, text width=4mm]
    
   \tikzstyle{act}=[	circle, thick, minimum size=0.4cm,
   				draw=black!85, fill=DanYellow!80, inner sep=0pt, text width=4mm]
    
   \tikzstyle{Imp}=[	circle, thick, minimum size=0.4cm, draw=black!85,
				fill=DanGreen!80, inner sep=0pt, text width=4mm]
				
   \tikzstyle{Bath}=[	circle, thick, minimum size=0.4cm, draw=black!85,
				fill=black!30, inner sep=0pt, text width=4mm]
    
    	\matrix (m2) [matrix of nodes, column sep=10, row sep = 8] {
       		|[Bath]| & |[Bath]| & |[Imp]| & |[Imp]| & |[Bath]| & |[Bath]|  \\
    	};
    
    	\foreach \x/\y in {1/2,2/3,3/4,4/5,5/6} \draw (m2-1-\x) -- (m2-1-\y);
	\foreach \x in {1,2,3,4,5,6} \draw (m2-1-\x) -- ($(m2-1-\x)+(0,0.35)$);

	\draw [ 	thick, decoration={ brace, mirror, raise=0.5cm },
    			decorate]  (m2-1-1.west) -- (m2-1-2.east) %
		node [pos=0.5,anchor=north,yshift=-0.55cm] {$\downarrow$-Bath}; 
	\draw [ 	thick, decoration={ brace, mirror, raise=0.5cm },
    			decorate]  (m2-1-3.west) -- (m2-1-4.east) %
		node [pos=0.5,anchor=north,yshift=-0.55cm] {Impurity}; 
		
	\draw [ 	thick, decoration={ brace, mirror, raise=0.5cm },
    			decorate]  (m2-1-5.west) -- (m2-1-6.east) %
		node [pos=0.5,anchor=north,yshift=-0.55cm] {$\uparrow$-Bath};

\end{tikzpicture}
   \caption{
      Depiction of the MPS tensor network used to solve impurity models Eq.~(\ref{eq:Himp}). The physical Hilbert space dimensions of the sites are either $4$ if purified or $2$ if not purified. }
  \label{fig:MPS}
\end{figure}

The MPS used in this work is depicted in Fig.~\ref{fig:MPS}. The impurity degrees of freedom are at the center of the MPS with the spin-up and spin-down bath sorted according to their energy $\epsilon_k$ to the left/right of the impurity. The long-range couplings $V_k$ of the Hamiltonian Eq.~\ref{eq:Himp} are no issue for TDVP since the matrix product operator can be computed efficiently with links of dimension $5$ only. For tDMRG, we use swap-gates as discussed next.

\subsection*{tDMRG - Partial Swap Gates}
The tDMRG-like time evolution discussed in Ref.~\cite{FTPS} uses swap-gates to efficiently encode the long-range hybridization terms of strength $V_k$. In case we swap a purified site with a normal site, we choose to swap the physical degree of freedom only and keep the auxiliary one untouched, to keep the local degrees of freedom the same size at all times.

We begin with the usual swap gates for fermionic degrees of freedom:
\begin{equation*}
    S_{dd} = \sum_{n_i, n_j} (-1)^{n_i n_j} |n_i n_j\rangle \langle n_j n_i|,
\end{equation*}
where $\ket{i}$ and $\ket{j}$ are the basis states of the two sites $i$ and $j$ respectively and $n_i$ and $n_j$ are the total occupations. The factor $(-1)^{n_i n_j}$ gives a negative sign if both of the occupations are odd and accounts for the change of fermionic order caused by the swap. These swap gates are used if both Hilbert space dimensions are equal as indicated by the subscript $dd$, meaning that such a gate is used to swap two normal sites as well as two-purified sites.

To swap a normal site with a purified site, the order in which these two sites appear in the fermionic order matters. The fermionic order on the purified sites is chosen as: $ \ket{n_P n_A} = (c_{P}^\dagger)^{n_{P}} (c_{A}^\dagger)^{n_{A}} \ket{0}$ where $P$ ($A$) refers to the physical (auxiliary) degree of freedom respectively. So if the normal site comes before the purified site, the swap gate is essentially a generalization of the swap gate above. The auxiliary site is inactive because it is to the right of the two degrees of freedom being swapped: 
\begin{equation*}
S_{24} = S_{22} \otimes \mathbb{1},
\end{equation*}
where $\mathbb{1}$ is the $2 \times 2$ identity matrix acting on the auxiliary degree of freedom.

If, on the other hand, the normal site comes after the purified site in the fermionic order, the auxiliary site has an effect, because it is in between the two sites being swapped. The resulting swap gate is then given by:
\begin{equation*}
S_{42} =  \sum_{n_P n_A n_j }  (-1)^{n n_j + n_{P} n_{A}} |n_{P} n_{A} n_j\rangle \langle n_j n_{A} n_{P}|,
\end{equation*}
where $n = n_P + n_A$ is the total occupation of the purified site. The sign-factor can be understood as first swapping $n_j$ particles with all $n$ particles on the purified site and then swapping the physical- with the auxiliary degree of freedom of the purified site.

\subsection*{TDVP}
The TDVP time evolution is performed with a modified integration scheme compared to the usual right-left-right sweep~\cite{TDVP}. Instead, we start the integration in the middle at the impurity, sweep outwards one bath, jump back to the middle and sweep outwards the other bath. Details can be found in the Appendix of Ref~\cite{Bertrand_QQMC}. The reason for doing that is two-fold. First, the application of the creation/annihilation operators during the computation of the Green function destroys some tensor network basis states. If we start the time evolution at the sites where this happened allows TDVP to immediately recover the lost bond dimension and reduces the numerical error. The second reason is of practical nature: it is simply easier to generalize to multi-orbital impurity problems, where in the FTPS tensor network, there is no clear left and right anymore.

\section{Green Function Histogram}\label{app:histogram}
\begin{figure*}[t]
    \centering
    \includegraphics[width=1.7\columnwidth]{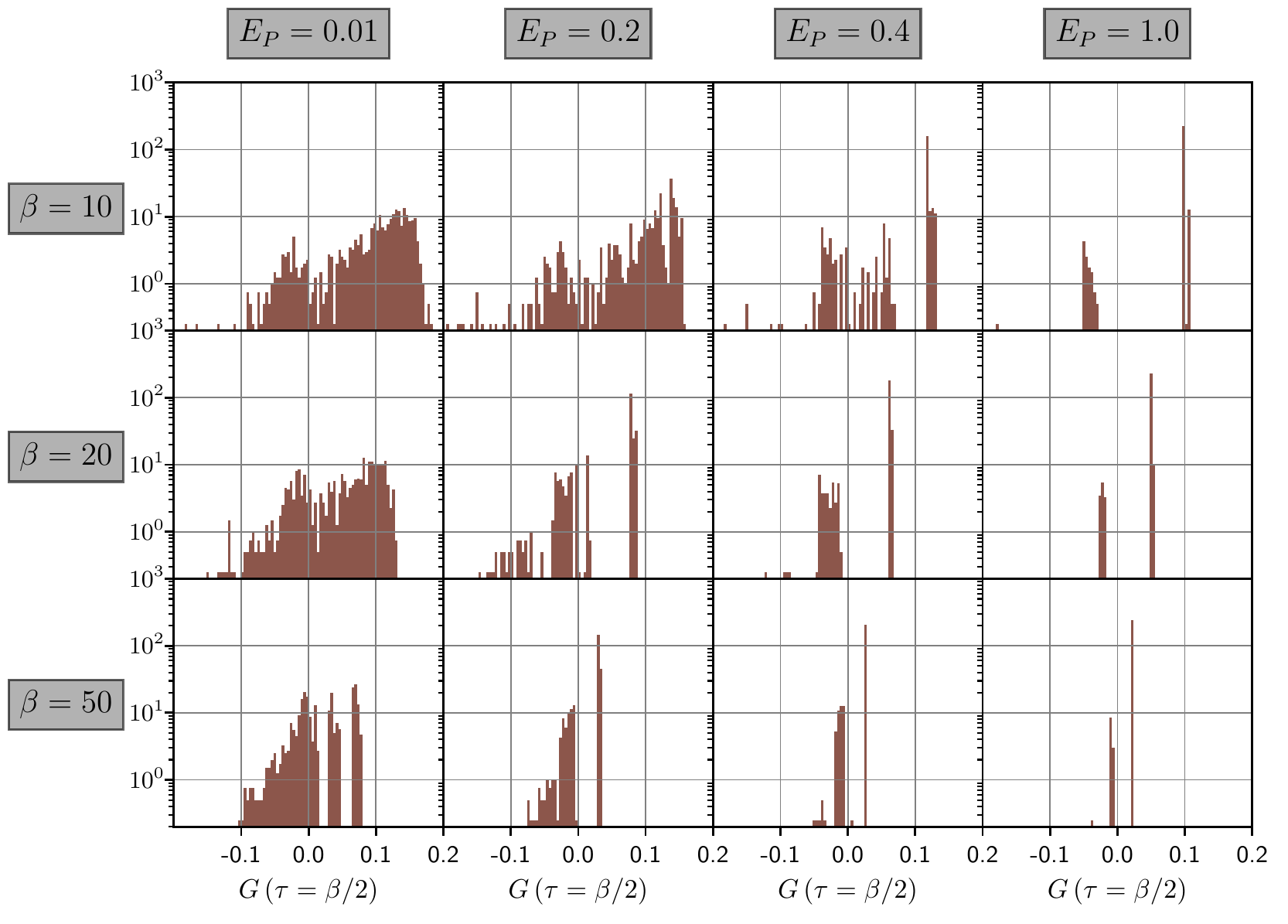}
    \caption{Histogram of single particle Green function at $\tau = \beta/2$ taken from samples of size $1000$ at $U=2$. The left-most column with $E_P = 0.01$ purifies only the impurity and the $\epsilon_k = 0 $ bath site. Decreasing $\beta$ as well as increasing $E_P$ reduces the total width of the histogram. Importantly, with increasing $E_P$, a single METTS starts to dominate the distribution (called highest probability METTS in the main text). }
     \label{fig:histogram}
\end{figure*}

As mentioned in the main text, the distribution of METTS Green functions is not Gaussian, especially for large values of $E_P$ and around $\beta/2$. Instead, single METTS dominates the distribution and a couple of side \emph{bands} appear as shown in Fig.~\ref{fig:histogram} (note the logarithmic scale on the y-axis). Even though sampling from such distribution is hard, the highest probability METTS alone often gives a very satisfying result already. Fortunately, in case that computing the highest probability METTS with a sufficiently high $E_P$ is computationally too expensive, decreasing $E_P$ improves the distribution and makes sampling easier.

\section{Additional Results}\label{app:additionalresults}

\begin{figure}[ht!]
    %\centering
    \includegraphics[width=0.9\columnwidth]{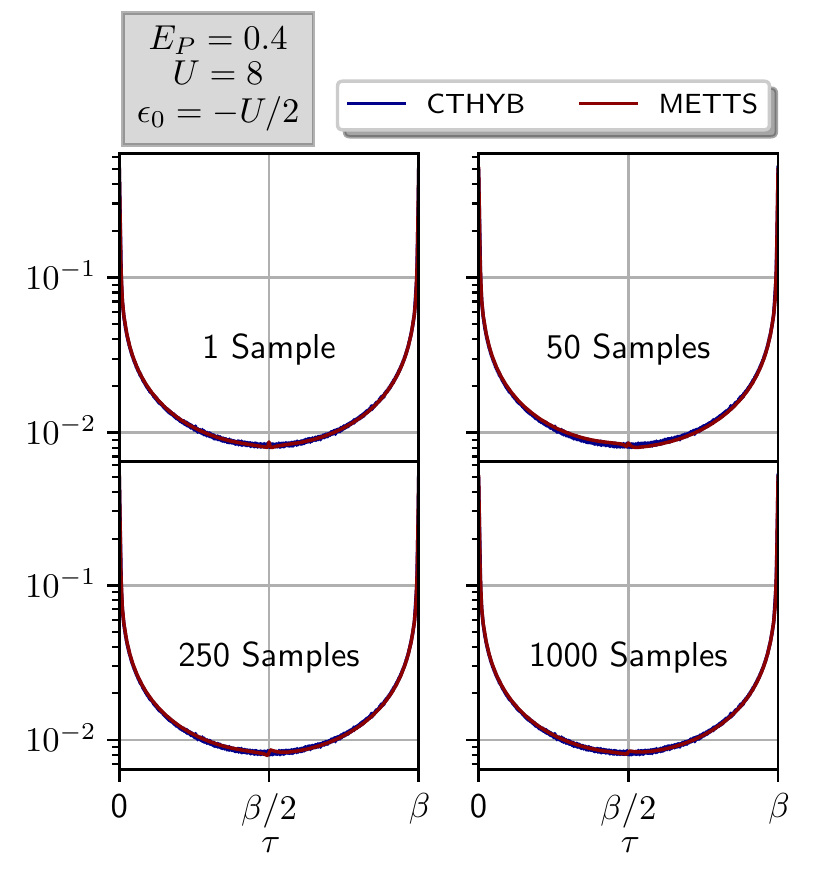}
    \caption{Convergence plot similar to Fig.~\ref{fig:U2Convergence} with very strong interaction $U=8$ and $\beta=100$. The result is comparable to its $U=2$ counterpart and the agreement to CTHYB is even better with nearly negligible errors already for the highest probability METTS. }
    \label{fig:U8Convergence}
\end{figure}

\begin{figure}[ht!]
    %\centering
    \includegraphics[width=0.9\columnwidth]{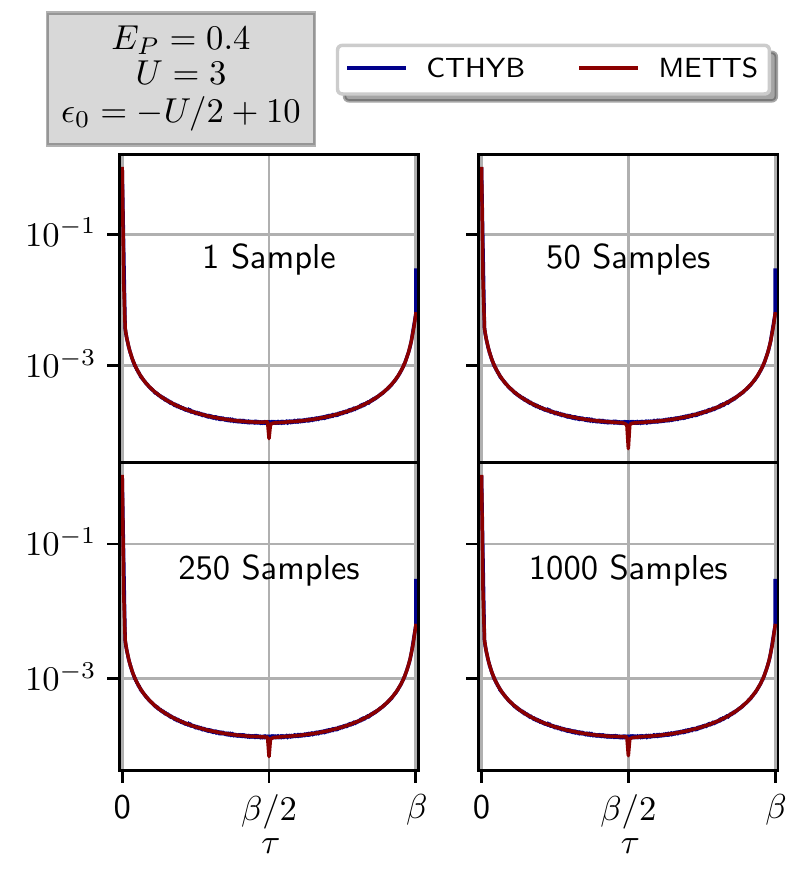}
    \caption{Convergence plot similar to Fig.~\ref{fig:U2Convergence} for a system at $\beta = 100$ with a strongly asymmetric on-site energy $\epsilon_0=-U/2 +10$ such that the impurity occupation is only $\langle n_\sigma \rangle \approx 0.006$. Again the agreement is very good especially considering that the Green function at $\beta/2$ is about 2 orders of magnitude smaller than those in Fig.~\ref{fig:U2Convergence} or Fig.~\ref{fig:U8Convergence}. The difference exactly at $\beta$ is attributed to CTHYB sampling problems caused by the low occupation. This means that the METTS result is the correct one at $\tau=\beta$. }
    \label{fig:U3AsymConvergence}
\end{figure}

To obtain a complete picture of the METTS sampling of Impurity Green functions, lets us study two additional systems, both in somewhat extreme parameter regions. In Fig.~\ref{fig:U8Convergence}, we show Green functions for very strong interactions $U=8$ and in Fig.~\ref{fig:U3AsymConvergence} for a strongly asymmetric impurity model. Both cases actually perform better than the symmetric $U=2$ (Fig~\ref{fig:U2Convergence}) counterpart with tiny errors even when using just a single sample. This result exemplifies the wide range of applicability of this method with respect to different parameter regimes.

\bibliography{biblio}

\begin{thebibliography}{45}%
\makeatletter
\providecommand \@ifxundefined [1]{%
 \@ifx{#1\undefined}
}%
\providecommand \@ifnum [1]{%
 \ifnum #1\expandafter \@firstoftwo
 \else \expandafter \@secondoftwo
 \fi
}%
\providecommand \@ifx [1]{%
 \ifx #1\expandafter \@firstoftwo
 \else \expandafter \@secondoftwo
 \fi
}%
\providecommand \natexlab [1]{#1}%
\providecommand \enquote  [1]{``#1''}%
\providecommand \bibnamefont  [1]{#1}%
\providecommand \bibfnamefont [1]{#1}%
\providecommand \citenamefont [1]{#1}%
\providecommand \href@noop [0]{\@secondoftwo}%
\providecommand \href [0]{\begingroup \@sanitize@url \@href}%
\providecommand \@href[1]{\@@startlink{#1}\@@href}%
\providecommand \@@href[1]{\endgroup#1\@@endlink}%
\providecommand \@sanitize@url [0]{\catcode `\\12\catcode `\$12\catcode
  `\&12\catcode `\#12\catcode `\^12\catcode `\_12\catcode `\%12\relax}%
\providecommand \@@startlink[1]{}%
\providecommand \@@endlink[0]{}%
\providecommand \url  [0]{\begingroup\@sanitize@url \@url }%
\providecommand \@url [1]{\endgroup\@href {#1}{\urlprefix }}%
\providecommand \urlprefix  [0]{URL }%
\providecommand \Eprint [0]{\href }%
\providecommand \doibase [0]{https://doi.org/}%
\providecommand \selectlanguage [0]{\@gobble}%
\providecommand \bibinfo  [0]{\@secondoftwo}%
\providecommand \bibfield  [0]{\@secondoftwo}%
\providecommand \translation [1]{[#1]}%
\providecommand \BibitemOpen [0]{}%
\providecommand \bibitemStop [0]{}%
\providecommand \bibitemNoStop [0]{.\EOS\space}%
\providecommand \EOS [0]{\spacefactor3000\relax}%
\providecommand \BibitemShut  [1]{\csname bibitem#1\endcsname}%
\let\auto@bib@innerbib\@empty
%</preamble>
\bibitem [{\citenamefont {White}(1992)}]{White_DMRG}%
  \BibitemOpen
  \bibfield  {author} {\bibinfo {author} {\bibfnamefont {S.~R.}\ \bibnamefont
  {White}},\ }\bibfield  {title} {\bibinfo {title} {Density matrix formulation
  for quantum renormalization groups},\ }\href
  {https://doi.org/10.1103/PhysRevLett.69.2863} {\bibfield  {journal} {\bibinfo
   {journal} {Phys. Rev. Lett.}\ }\textbf {\bibinfo {volume} {69}},\ \bibinfo
  {pages} {2863} (\bibinfo {year} {1992})}\BibitemShut {NoStop}%
\bibitem [{\citenamefont {Schollw\"ock}(2005)}]{UliRevModPhys.77.259}%
  \BibitemOpen
  \bibfield  {author} {\bibinfo {author} {\bibfnamefont {U.}~\bibnamefont
  {Schollw\"ock}},\ }\bibfield  {title} {\bibinfo {title} {The density-matrix
  renormalization group},\ }\href {https://doi.org/10.1103/RevModPhys.77.259}
  {\bibfield  {journal} {\bibinfo  {journal} {Rev. Mod. Phys.}\ }\textbf
  {\bibinfo {volume} {77}},\ \bibinfo {pages} {259} (\bibinfo {year}
  {2005})}\BibitemShut {NoStop}%
\bibitem [{\citenamefont {Schollw\"ock}(2011)}]{SCHOLLWOCK201196}%
  \BibitemOpen
  \bibfield  {author} {\bibinfo {author} {\bibfnamefont {U.}~\bibnamefont
  {Schollw\"ock}},\ }\bibfield  {title} {\bibinfo {title} {The density-matrix
  renormalization group in the age of matrix product states},\ }\href
  {https://doi.org/https://doi.org/10.1016/j.aop.2010.09.012} {\bibfield
  {journal} {\bibinfo  {journal} {Annals of Physics}\ }\textbf {\bibinfo
  {volume} {326}},\ \bibinfo {pages} {96} (\bibinfo {year} {2011})},\ \bibinfo
  {note} {january 2011 Special Issue}\BibitemShut {NoStop}%
\bibitem [{\citenamefont {Paeckel}\ \emph {et~al.}(2019)\citenamefont
  {Paeckel}, \citenamefont {Köhler}, \citenamefont {Swoboda}, \citenamefont
  {Manmana}, \citenamefont {Schollwöck},\ and\ \citenamefont
  {Hubig}}]{Paeckel}%
  \BibitemOpen
  \bibfield  {author} {\bibinfo {author} {\bibfnamefont {S.}~\bibnamefont
  {Paeckel}}, \bibinfo {author} {\bibfnamefont {T.}~\bibnamefont {Köhler}},
  \bibinfo {author} {\bibfnamefont {A.}~\bibnamefont {Swoboda}}, \bibinfo
  {author} {\bibfnamefont {S.~R.}\ \bibnamefont {Manmana}}, \bibinfo {author}
  {\bibfnamefont {U.}~\bibnamefont {Schollwöck}},\ and\ \bibinfo {author}
  {\bibfnamefont {C.}~\bibnamefont {Hubig}},\ }\bibfield  {title} {\bibinfo
  {title} {Time-evolution methods for matrix-product states},\ }\href
  {https://doi.org/https://doi.org/10.1016/j.aop.2019.167998} {\bibfield
  {journal} {\bibinfo  {journal} {Annals of Physics}\ }\textbf {\bibinfo
  {volume} {411}},\ \bibinfo {pages} {167998} (\bibinfo {year}
  {2019})}\BibitemShut {NoStop}%
\bibitem [{\citenamefont {Cohen}\ \emph {et~al.}(2015)\citenamefont {Cohen},
  \citenamefont {Gull}, \citenamefont {Reichman},\ and\ \citenamefont
  {Millis}}]{InchwormCohenGullMillisRealTime2015}%
  \BibitemOpen
  \bibfield  {author} {\bibinfo {author} {\bibfnamefont {G.}~\bibnamefont
  {Cohen}}, \bibinfo {author} {\bibfnamefont {E.}~\bibnamefont {Gull}},
  \bibinfo {author} {\bibfnamefont {D.~R.}\ \bibnamefont {Reichman}},\ and\
  \bibinfo {author} {\bibfnamefont {A.~J.}\ \bibnamefont {Millis}},\ }\bibfield
   {title} {\bibinfo {title} {Taming the dynamical sign problem in real-time
  evolution of quantum many-body problems},\ }\href
  {https://doi.org/10.1103/PhysRevLett.115.266802} {\bibfield  {journal}
  {\bibinfo  {journal} {Phys. Rev. Lett.}\ }\textbf {\bibinfo {volume} {115}},\
  \bibinfo {pages} {266802} (\bibinfo {year} {2015})}\BibitemShut {NoStop}%
\bibitem [{\citenamefont {Profumo}\ \emph {et~al.}(2015)\citenamefont
  {Profumo}, \citenamefont {Groth}, \citenamefont {Messio}, \citenamefont
  {Parcollet},\ and\ \citenamefont {Waintal}}]{ProfumoRealTimeQMC}%
  \BibitemOpen
  \bibfield  {author} {\bibinfo {author} {\bibfnamefont {R.~E.~V.}\
  \bibnamefont {Profumo}}, \bibinfo {author} {\bibfnamefont {C.}~\bibnamefont
  {Groth}}, \bibinfo {author} {\bibfnamefont {L.}~\bibnamefont {Messio}},
  \bibinfo {author} {\bibfnamefont {O.}~\bibnamefont {Parcollet}},\ and\
  \bibinfo {author} {\bibfnamefont {X.}~\bibnamefont {Waintal}},\ }\bibfield
  {title} {\bibinfo {title} {Quantum monte carlo for correlated
  out-of-equilibrium nanoelectronic devices},\ }\href
  {https://doi.org/10.1103/PhysRevB.91.245154} {\bibfield  {journal} {\bibinfo
  {journal} {Phys. Rev. B}\ }\textbf {\bibinfo {volume} {91}},\ \bibinfo
  {pages} {245154} (\bibinfo {year} {2015})}\BibitemShut {NoStop}%
\bibitem [{\citenamefont {Dong}\ \emph {et~al.}(2017)\citenamefont {Dong},
  \citenamefont {Krivenko}, \citenamefont {Kleinhenz}, \citenamefont {Antipov},
  \citenamefont {Cohen},\ and\ \citenamefont {Gull}}]{Gull_RealTimeMC}%
  \BibitemOpen
  \bibfield  {author} {\bibinfo {author} {\bibfnamefont {Q.}~\bibnamefont
  {Dong}}, \bibinfo {author} {\bibfnamefont {I.}~\bibnamefont {Krivenko}},
  \bibinfo {author} {\bibfnamefont {J.}~\bibnamefont {Kleinhenz}}, \bibinfo
  {author} {\bibfnamefont {A.~E.}\ \bibnamefont {Antipov}}, \bibinfo {author}
  {\bibfnamefont {G.}~\bibnamefont {Cohen}},\ and\ \bibinfo {author}
  {\bibfnamefont {E.}~\bibnamefont {Gull}},\ }\bibfield  {title} {\bibinfo
  {title} {Quantum monte carlo solution of the dynamical mean field equations
  in real time},\ }\href {https://doi.org/10.1103/PhysRevB.96.155126}
  {\bibfield  {journal} {\bibinfo  {journal} {Phys. Rev. B}\ }\textbf {\bibinfo
  {volume} {96}},\ \bibinfo {pages} {155126} (\bibinfo {year}
  {2017})}\BibitemShut {NoStop}%
\bibitem [{\citenamefont {Bertrand}\ \emph {et~al.}(2019)\citenamefont
  {Bertrand}, \citenamefont {Florens}, \citenamefont {Parcollet},\ and\
  \citenamefont {Waintal}}]{BertrandPRX}%
  \BibitemOpen
  \bibfield  {author} {\bibinfo {author} {\bibfnamefont {C.}~\bibnamefont
  {Bertrand}}, \bibinfo {author} {\bibfnamefont {S.}~\bibnamefont {Florens}},
  \bibinfo {author} {\bibfnamefont {O.}~\bibnamefont {Parcollet}},\ and\
  \bibinfo {author} {\bibfnamefont {X.}~\bibnamefont {Waintal}},\ }\bibfield
  {title} {\bibinfo {title} {Reconstructing nonequilibrium regimes of quantum
  many-body systems from the analytical structure of perturbative expansions},\
  }\href {https://doi.org/10.1103/PhysRevX.9.041008} {\bibfield  {journal}
  {\bibinfo  {journal} {Phys. Rev. X}\ }\textbf {\bibinfo {volume} {9}},\
  \bibinfo {pages} {041008} (\bibinfo {year} {2019})}\BibitemShut {NoStop}%
\bibitem [{\citenamefont {Ma\ifmmode~\check{c}\else \v{c}\fi{}ek}\ \emph
  {et~al.}(2020)\citenamefont {Ma\ifmmode~\check{c}\else \v{c}\fi{}ek},
  \citenamefont {Dumitrescu}, \citenamefont {Bertrand}, \citenamefont {Triggs},
  \citenamefont {Parcollet},\ and\ \citenamefont {Waintal}}]{QQMCPRL}%
  \BibitemOpen
  \bibfield  {author} {\bibinfo {author} {\bibfnamefont {M.}~\bibnamefont
  {Ma\ifmmode~\check{c}\else \v{c}\fi{}ek}}, \bibinfo {author} {\bibfnamefont
  {P.~T.}\ \bibnamefont {Dumitrescu}}, \bibinfo {author} {\bibfnamefont
  {C.}~\bibnamefont {Bertrand}}, \bibinfo {author} {\bibfnamefont
  {B.}~\bibnamefont {Triggs}}, \bibinfo {author} {\bibfnamefont
  {O.}~\bibnamefont {Parcollet}},\ and\ \bibinfo {author} {\bibfnamefont
  {X.}~\bibnamefont {Waintal}},\ }\bibfield  {title} {\bibinfo {title} {Quantum
  quasi-monte carlo technique for many-body perturbative expansions},\ }\href
  {https://doi.org/10.1103/PhysRevLett.125.047702} {\bibfield  {journal}
  {\bibinfo  {journal} {Phys. Rev. Lett.}\ }\textbf {\bibinfo {volume} {125}},\
  \bibinfo {pages} {047702} (\bibinfo {year} {2020})}\BibitemShut {NoStop}%
\bibitem [{\citenamefont {Prokof'ev}\ and\ \citenamefont
  {Svistunov}(1998)}]{ProkofievSvistunovDiagQMCPhysRevLett.81.2514}%
  \BibitemOpen
  \bibfield  {author} {\bibinfo {author} {\bibfnamefont {N.~V.}\ \bibnamefont
  {Prokof'ev}}\ and\ \bibinfo {author} {\bibfnamefont {B.~V.}\ \bibnamefont
  {Svistunov}},\ }\bibfield  {title} {\bibinfo {title} {Polaron problem by
  diagrammatic quantum monte carlo},\ }\href
  {https://doi.org/10.1103/PhysRevLett.81.2514} {\bibfield  {journal} {\bibinfo
   {journal} {Phys. Rev. Lett.}\ }\textbf {\bibinfo {volume} {81}},\ \bibinfo
  {pages} {2514} (\bibinfo {year} {1998})}\BibitemShut {NoStop}%
\bibitem [{\citenamefont {White}(2009)}]{White_Metts}%
  \BibitemOpen
  \bibfield  {author} {\bibinfo {author} {\bibfnamefont {S.~R.}\ \bibnamefont
  {White}},\ }\bibfield  {title} {\bibinfo {title} {Minimally entangled typical
  quantum states at finite temperature},\ }\href
  {https://doi.org/10.1103/PhysRevLett.102.190601} {\bibfield  {journal}
  {\bibinfo  {journal} {Phys. Rev. Lett.}\ }\textbf {\bibinfo {volume} {102}},\
  \bibinfo {pages} {190601} (\bibinfo {year} {2009})}\BibitemShut {NoStop}%
\bibitem [{\citenamefont {Stoudenmire}\ and\ \citenamefont
  {White}(2010)}]{Miles_Metts}%
  \BibitemOpen
  \bibfield  {author} {\bibinfo {author} {\bibfnamefont {E.~M.}\ \bibnamefont
  {Stoudenmire}}\ and\ \bibinfo {author} {\bibfnamefont {S.~R.}\ \bibnamefont
  {White}},\ }\bibfield  {title} {\bibinfo {title} {Minimally entangled typical
  thermal state algorithms},\ }\href
  {https://doi.org/10.1088/1367-2630/12/5/055026} {\bibfield  {journal}
  {\bibinfo  {journal} {New Journal of Physics}\ }\textbf {\bibinfo {volume}
  {12}},\ \bibinfo {pages} {055026} (\bibinfo {year} {2010})}\BibitemShut
  {NoStop}%
\bibitem [{\citenamefont {Binder}\ and\ \citenamefont
  {Barthel}(2017)}]{Barthel_SymmetricMetts}%
  \BibitemOpen
  \bibfield  {author} {\bibinfo {author} {\bibfnamefont {M.}~\bibnamefont
  {Binder}}\ and\ \bibinfo {author} {\bibfnamefont {T.}~\bibnamefont
  {Barthel}},\ }\bibfield  {title} {\bibinfo {title} {Symmetric minimally
  entangled typical thermal states for canonical and grand-canonical
  ensembles},\ }\href {https://doi.org/10.1103/PhysRevB.95.195148} {\bibfield
  {journal} {\bibinfo  {journal} {Phys. Rev. B}\ }\textbf {\bibinfo {volume}
  {95}},\ \bibinfo {pages} {195148} (\bibinfo {year} {2017})}\BibitemShut
  {NoStop}%
\bibitem [{\citenamefont {Chen}\ and\ \citenamefont
  {Stoudenmire}(2020)}]{Jing_HybridPurificationMetts}%
  \BibitemOpen
  \bibfield  {author} {\bibinfo {author} {\bibfnamefont {J.}~\bibnamefont
  {Chen}}\ and\ \bibinfo {author} {\bibfnamefont {E.~M.}\ \bibnamefont
  {Stoudenmire}},\ }\bibfield  {title} {\bibinfo {title} {Hybrid purification
  and sampling approach for thermal quantum systems},\ }\href
  {https://doi.org/10.1103/PhysRevB.101.195119} {\bibfield  {journal} {\bibinfo
   {journal} {Phys. Rev. B}\ }\textbf {\bibinfo {volume} {101}},\ \bibinfo
  {pages} {195119} (\bibinfo {year} {2020})}\BibitemShut {NoStop}%
\bibitem [{\citenamefont {Chung}\ and\ \citenamefont
  {Schollw\"ock}(2019)}]{chung_2019minimally}%
  \BibitemOpen
  \bibfield  {author} {\bibinfo {author} {\bibfnamefont {C.-M.}\ \bibnamefont
  {Chung}}\ and\ \bibinfo {author} {\bibfnamefont {U.}~\bibnamefont
  {Schollw\"ock}},\ }\bibfield  {title} {\bibinfo {title} {Minimally entangled
  typical thermal states with auxiliary matrix-product-state bases},\
  }\href@noop {} {\bibfield  {journal} {\bibinfo  {journal} {arxiv}\ }
  (\bibinfo {year} {2019})},\ \Eprint {https://arxiv.org/abs/1910.03329}
  {arXiv:1910.03329 [cond-mat.str-el]} \BibitemShut {NoStop}%
\bibitem [{\citenamefont {Feiguin}\ and\ \citenamefont
  {White}(2005)}]{FeiguinPurification}%
  \BibitemOpen
  \bibfield  {author} {\bibinfo {author} {\bibfnamefont {A.~E.}\ \bibnamefont
  {Feiguin}}\ and\ \bibinfo {author} {\bibfnamefont {S.~R.}\ \bibnamefont
  {White}},\ }\bibfield  {title} {\bibinfo {title} {Finite-temperature density
  matrix renormalization using an enlarged hilbert space},\ }\href
  {https://doi.org/10.1103/PhysRevB.72.220401} {\bibfield  {journal} {\bibinfo
  {journal} {Phys. Rev. B}\ }\textbf {\bibinfo {volume} {72}},\ \bibinfo
  {pages} {220401} (\bibinfo {year} {2005})}\BibitemShut {NoStop}%
\bibitem [{\citenamefont {Zwolak}\ and\ \citenamefont {Vidal}(2004)}]{Zwolak}%
  \BibitemOpen
  \bibfield  {author} {\bibinfo {author} {\bibfnamefont {M.}~\bibnamefont
  {Zwolak}}\ and\ \bibinfo {author} {\bibfnamefont {G.}~\bibnamefont {Vidal}},\
  }\bibfield  {title} {\bibinfo {title} {Mixed-state dynamics in
  one-dimensional quantum lattice systems: A time-dependent superoperator
  renormalization algorithm},\ }\href
  {https://doi.org/10.1103/PhysRevLett.93.207205} {\bibfield  {journal}
  {\bibinfo  {journal} {Phys. Rev. Lett.}\ }\textbf {\bibinfo {volume} {93}},\
  \bibinfo {pages} {207205} (\bibinfo {year} {2004})}\BibitemShut {NoStop}%
\bibitem [{\citenamefont {Verstraete}\ \emph {et~al.}(2004)\citenamefont
  {Verstraete}, \citenamefont {Garc\'{\i}a-Ripoll},\ and\ \citenamefont
  {Cirac}}]{Verstraete_purification}%
  \BibitemOpen
  \bibfield  {author} {\bibinfo {author} {\bibfnamefont {F.}~\bibnamefont
  {Verstraete}}, \bibinfo {author} {\bibfnamefont {J.~J.}\ \bibnamefont
  {Garc\'{\i}a-Ripoll}},\ and\ \bibinfo {author} {\bibfnamefont {J.~I.}\
  \bibnamefont {Cirac}},\ }\bibfield  {title} {\bibinfo {title} {Matrix product
  density operators: Simulation of finite-temperature and dissipative
  systems},\ }\href {https://doi.org/10.1103/PhysRevLett.93.207204} {\bibfield
  {journal} {\bibinfo  {journal} {Phys. Rev. Lett.}\ }\textbf {\bibinfo
  {volume} {93}},\ \bibinfo {pages} {207204} (\bibinfo {year}
  {2004})}\BibitemShut {NoStop}%
\bibitem [{\citenamefont {Hauschild}\ \emph {et~al.}(2018)\citenamefont
  {Hauschild}, \citenamefont {Leviatan}, \citenamefont {Bardarson},
  \citenamefont {Altman}, \citenamefont {Zaletel},\ and\ \citenamefont
  {Pollmann}}]{HauschildPurification}%
  \BibitemOpen
  \bibfield  {author} {\bibinfo {author} {\bibfnamefont {J.}~\bibnamefont
  {Hauschild}}, \bibinfo {author} {\bibfnamefont {E.}~\bibnamefont {Leviatan}},
  \bibinfo {author} {\bibfnamefont {J.~H.}\ \bibnamefont {Bardarson}}, \bibinfo
  {author} {\bibfnamefont {E.}~\bibnamefont {Altman}}, \bibinfo {author}
  {\bibfnamefont {M.~P.}\ \bibnamefont {Zaletel}},\ and\ \bibinfo {author}
  {\bibfnamefont {F.}~\bibnamefont {Pollmann}},\ }\bibfield  {title} {\bibinfo
  {title} {Finding purifications with minimal entanglement},\ }\href
  {https://doi.org/10.1103/PhysRevB.98.235163} {\bibfield  {journal} {\bibinfo
  {journal} {Phys. Rev. B}\ }\textbf {\bibinfo {volume} {98}},\ \bibinfo
  {pages} {235163} (\bibinfo {year} {2018})}\BibitemShut {NoStop}%
\bibitem [{\citenamefont {Georges}\ \emph {et~al.}(1996)\citenamefont
  {Georges}, \citenamefont {Kotliar}, \citenamefont {Krauth},\ and\
  \citenamefont {Rozenberg}}]{Georges_DMFT}%
  \BibitemOpen
  \bibfield  {author} {\bibinfo {author} {\bibfnamefont {A.}~\bibnamefont
  {Georges}}, \bibinfo {author} {\bibfnamefont {G.}~\bibnamefont {Kotliar}},
  \bibinfo {author} {\bibfnamefont {W.}~\bibnamefont {Krauth}},\ and\ \bibinfo
  {author} {\bibfnamefont {M.~J.}\ \bibnamefont {Rozenberg}},\ }\bibfield
  {title} {\bibinfo {title} {Dynamical mean-field theory of strongly correlated
  fermion systems and the limit of infinite dimensions},\ }\href
  {https://doi.org/10.1103/RevModPhys.68.13} {\bibfield  {journal} {\bibinfo
  {journal} {Rev. Mod. Phys.}\ }\textbf {\bibinfo {volume} {68}},\ \bibinfo
  {pages} {13} (\bibinfo {year} {1996})}\BibitemShut {NoStop}%
\bibitem [{\citenamefont {Kotliar}\ \emph {et~al.}(2006)\citenamefont
  {Kotliar}, \citenamefont {Savrasov}, \citenamefont {Haule}, \citenamefont
  {Oudovenko}, \citenamefont {Parcollet},\ and\ \citenamefont
  {Marianetti}}]{KotliarRMP2006}%
  \BibitemOpen
  \bibfield  {author} {\bibinfo {author} {\bibfnamefont {G.}~\bibnamefont
  {Kotliar}}, \bibinfo {author} {\bibfnamefont {S.~Y.}\ \bibnamefont
  {Savrasov}}, \bibinfo {author} {\bibfnamefont {K.}~\bibnamefont {Haule}},
  \bibinfo {author} {\bibfnamefont {V.~S.}\ \bibnamefont {Oudovenko}}, \bibinfo
  {author} {\bibfnamefont {O.}~\bibnamefont {Parcollet}},\ and\ \bibinfo
  {author} {\bibfnamefont {C.~A.}\ \bibnamefont {Marianetti}},\ }\bibfield
  {title} {\bibinfo {title} {Electronic structure calculations with dynamical
  mean-field theory},\ }\href {https://doi.org/10.1103/RevModPhys.78.865}
  {\bibfield  {journal} {\bibinfo  {journal} {Rev. Mod. Phys.}\ }\textbf
  {\bibinfo {volume} {78}},\ \bibinfo {pages} {865} (\bibinfo {year}
  {2006})}\BibitemShut {NoStop}%
\bibitem [{\citenamefont {Rubtsov}\ \emph {et~al.}(2005)\citenamefont
  {Rubtsov}, \citenamefont {Savkin},\ and\ \citenamefont
  {Lichtenstein}}]{RubtsovCTQMC2005}%
  \BibitemOpen
  \bibfield  {author} {\bibinfo {author} {\bibfnamefont {A.~N.}\ \bibnamefont
  {Rubtsov}}, \bibinfo {author} {\bibfnamefont {V.~V.}\ \bibnamefont
  {Savkin}},\ and\ \bibinfo {author} {\bibfnamefont {A.~I.}\ \bibnamefont
  {Lichtenstein}},\ }\bibfield  {title} {\bibinfo {title} {Continuous-time
  quantum monte carlo method for fermions},\ }\href
  {https://doi.org/10.1103/PhysRevB.72.035122} {\bibfield  {journal} {\bibinfo
  {journal} {Phys. Rev. B}\ }\textbf {\bibinfo {volume} {72}},\ \bibinfo
  {pages} {035122} (\bibinfo {year} {2005})}\BibitemShut {NoStop}%
\bibitem [{\citenamefont {Werner}\ \emph {et~al.}(2006)\citenamefont {Werner},
  \citenamefont {Comanac}, \citenamefont {de' Medici}, \citenamefont {Troyer},\
  and\ \citenamefont {Millis}}]{cthyb2006}%
  \BibitemOpen
  \bibfield  {author} {\bibinfo {author} {\bibfnamefont {P.}~\bibnamefont
  {Werner}}, \bibinfo {author} {\bibfnamefont {A.}~\bibnamefont {Comanac}},
  \bibinfo {author} {\bibfnamefont {L.}~\bibnamefont {de' Medici}}, \bibinfo
  {author} {\bibfnamefont {M.}~\bibnamefont {Troyer}},\ and\ \bibinfo {author}
  {\bibfnamefont {A.~J.}\ \bibnamefont {Millis}},\ }\bibfield  {title}
  {\bibinfo {title} {Continuous-time solver for quantum impurity models},\
  }\href {https://doi.org/10.1103/PhysRevLett.97.076405} {\bibfield  {journal}
  {\bibinfo  {journal} {Phys. Rev. Lett.}\ }\textbf {\bibinfo {volume} {97}},\
  \bibinfo {pages} {076405} (\bibinfo {year} {2006})}\BibitemShut {NoStop}%
\bibitem [{\citenamefont {Gull}\ \emph {et~al.}(2008)\citenamefont {Gull},
  \citenamefont {Werner}, \citenamefont {Parcollet},\ and\ \citenamefont
  {Troyer}}]{Gull_2008}%
  \BibitemOpen
  \bibfield  {author} {\bibinfo {author} {\bibfnamefont {E.}~\bibnamefont
  {Gull}}, \bibinfo {author} {\bibfnamefont {P.}~\bibnamefont {Werner}},
  \bibinfo {author} {\bibfnamefont {O.}~\bibnamefont {Parcollet}},\ and\
  \bibinfo {author} {\bibfnamefont {M.}~\bibnamefont {Troyer}},\ }\bibfield
  {title} {\bibinfo {title} {Continuous-time auxiliary-field monte carlo for
  quantum impurity models},\ }\href
  {https://doi.org/10.1209/0295-5075/82/57003} {\bibfield  {journal} {\bibinfo
  {journal} {EPL (Europhysics Letters)}\ }\textbf {\bibinfo {volume} {82}},\
  \bibinfo {pages} {57003} (\bibinfo {year} {2008})}\BibitemShut {NoStop}%
\bibitem [{\citenamefont {Gull}\ \emph {et~al.}(2011)\citenamefont {Gull},
  \citenamefont {Millis}, \citenamefont {Lichtenstein}, \citenamefont
  {Rubtsov}, \citenamefont {Troyer},\ and\ \citenamefont {Werner}}]{CTQMC_RMP}%
  \BibitemOpen
  \bibfield  {author} {\bibinfo {author} {\bibfnamefont {E.}~\bibnamefont
  {Gull}}, \bibinfo {author} {\bibfnamefont {A.~J.}\ \bibnamefont {Millis}},
  \bibinfo {author} {\bibfnamefont {A.~I.}\ \bibnamefont {Lichtenstein}},
  \bibinfo {author} {\bibfnamefont {A.~N.}\ \bibnamefont {Rubtsov}}, \bibinfo
  {author} {\bibfnamefont {M.}~\bibnamefont {Troyer}},\ and\ \bibinfo {author}
  {\bibfnamefont {P.}~\bibnamefont {Werner}},\ }\bibfield  {title} {\bibinfo
  {title} {Continuous-time monte carlo methods for quantum impurity models},\
  }\href {https://doi.org/10.1103/RevModPhys.83.349} {\bibfield  {journal}
  {\bibinfo  {journal} {Rev. Mod. Phys.}\ }\textbf {\bibinfo {volume} {83}},\
  \bibinfo {pages} {349} (\bibinfo {year} {2011})}\BibitemShut {NoStop}%
\bibitem [{\citenamefont {Bulla}\ \emph {et~al.}(2008)\citenamefont {Bulla},
  \citenamefont {Costi},\ and\ \citenamefont {Pruschke}}]{NRG_bulla}%
  \BibitemOpen
  \bibfield  {author} {\bibinfo {author} {\bibfnamefont {R.}~\bibnamefont
  {Bulla}}, \bibinfo {author} {\bibfnamefont {T.~A.}\ \bibnamefont {Costi}},\
  and\ \bibinfo {author} {\bibfnamefont {T.}~\bibnamefont {Pruschke}},\
  }\bibfield  {title} {\bibinfo {title} {Numerical renormalization group method
  for quantum impurity systems},\ }\href
  {https://doi.org/10.1103/RevModPhys.80.395} {\bibfield  {journal} {\bibinfo
  {journal} {Rev. Mod. Phys.}\ }\textbf {\bibinfo {volume} {80}},\ \bibinfo
  {pages} {395} (\bibinfo {year} {2008})}\BibitemShut {NoStop}%
\bibitem [{\citenamefont {Wilson}(1975)}]{NRG_original}%
  \BibitemOpen
  \bibfield  {author} {\bibinfo {author} {\bibfnamefont {K.~G.}\ \bibnamefont
  {Wilson}},\ }\bibfield  {title} {\bibinfo {title} {The renormalization group:
  Critical phenomena and the kondo problem},\ }\href
  {https://doi.org/10.1103/RevModPhys.47.773} {\bibfield  {journal} {\bibinfo
  {journal} {Rev. Mod. Phys.}\ }\textbf {\bibinfo {volume} {47}},\ \bibinfo
  {pages} {773} (\bibinfo {year} {1975})}\BibitemShut {NoStop}%
\bibitem [{\citenamefont {Jeckelmann}(2002)}]{JeckelmannPhysRevB.66.045114}%
  \BibitemOpen
  \bibfield  {author} {\bibinfo {author} {\bibfnamefont {E.}~\bibnamefont
  {Jeckelmann}},\ }\bibfield  {title} {\bibinfo {title} {Dynamical
  density-matrix renormalization-group method},\ }\href
  {https://doi.org/10.1103/PhysRevB.66.045114} {\bibfield  {journal} {\bibinfo
  {journal} {Phys. Rev. B}\ }\textbf {\bibinfo {volume} {66}},\ \bibinfo
  {pages} {045114} (\bibinfo {year} {2002})}\BibitemShut {NoStop}%
\bibitem [{\citenamefont {Garc\'{\i}a}\ \emph {et~al.}(2004)\citenamefont
  {Garc\'{\i}a}, \citenamefont {Hallberg},\ and\ \citenamefont
  {Rozenberg}}]{PhysRevLett.93.246403}%
  \BibitemOpen
  \bibfield  {author} {\bibinfo {author} {\bibfnamefont {D.~J.}\ \bibnamefont
  {Garc\'{\i}a}}, \bibinfo {author} {\bibfnamefont {K.}~\bibnamefont
  {Hallberg}},\ and\ \bibinfo {author} {\bibfnamefont {M.~J.}\ \bibnamefont
  {Rozenberg}},\ }\bibfield  {title} {\bibinfo {title} {Dynamical mean field
  theory with the density matrix renormalization group},\ }\href
  {https://doi.org/10.1103/PhysRevLett.93.246403} {\bibfield  {journal}
  {\bibinfo  {journal} {Phys. Rev. Lett.}\ }\textbf {\bibinfo {volume} {93}},\
  \bibinfo {pages} {246403} (\bibinfo {year} {2004})}\BibitemShut {NoStop}%
\bibitem [{\citenamefont {Karski}\ \emph {et~al.}(2005)\citenamefont {Karski},
  \citenamefont {Raas},\ and\ \citenamefont {Uhrig}}]{PhysRevB.72.113110}%
  \BibitemOpen
  \bibfield  {author} {\bibinfo {author} {\bibfnamefont {M.}~\bibnamefont
  {Karski}}, \bibinfo {author} {\bibfnamefont {C.}~\bibnamefont {Raas}},\ and\
  \bibinfo {author} {\bibfnamefont {G.~S.}\ \bibnamefont {Uhrig}},\ }\bibfield
  {title} {\bibinfo {title} {Electron spectra close to a metal-to-insulator
  transition},\ }\href {https://doi.org/10.1103/PhysRevB.72.113110} {\bibfield
  {journal} {\bibinfo  {journal} {Phys. Rev. B}\ }\textbf {\bibinfo {volume}
  {72}},\ \bibinfo {pages} {113110} (\bibinfo {year} {2005})}\BibitemShut
  {NoStop}%
\bibitem [{\citenamefont {Hallberg}\ \emph {et~al.}(1995)\citenamefont
  {Hallberg}, \citenamefont {Horsch},\ and\ \citenamefont
  {Mart\'{\i}nez}}]{PhysRevB.52.R719}%
  \BibitemOpen
  \bibfield  {author} {\bibinfo {author} {\bibfnamefont {K.~A.}\ \bibnamefont
  {Hallberg}}, \bibinfo {author} {\bibfnamefont {P.}~\bibnamefont {Horsch}},\
  and\ \bibinfo {author} {\bibfnamefont {G.}~\bibnamefont {Mart\'{\i}nez}},\
  }\bibfield  {title} {\bibinfo {title} {Numerical renormalization-group study
  of the correlation functions of the antiferromagnetic spin-1/2 heisenberg
  chain},\ }\href {https://doi.org/10.1103/PhysRevB.52.R719} {\bibfield
  {journal} {\bibinfo  {journal} {Phys. Rev. B}\ }\textbf {\bibinfo {volume}
  {52}},\ \bibinfo {pages} {R719} (\bibinfo {year} {1995})}\BibitemShut
  {NoStop}%
\bibitem [{\citenamefont {Wolf}\ \emph
  {et~al.}(2015{\natexlab{a}})\citenamefont {Wolf}, \citenamefont {Go},
  \citenamefont {McCulloch}, \citenamefont {Millis},\ and\ \citenamefont
  {Schollw\"ock}}]{Schollwoeck_ImagTimeSolver}%
  \BibitemOpen
  \bibfield  {author} {\bibinfo {author} {\bibfnamefont {F.~A.}\ \bibnamefont
  {Wolf}}, \bibinfo {author} {\bibfnamefont {A.}~\bibnamefont {Go}}, \bibinfo
  {author} {\bibfnamefont {I.~P.}\ \bibnamefont {McCulloch}}, \bibinfo {author}
  {\bibfnamefont {A.~J.}\ \bibnamefont {Millis}},\ and\ \bibinfo {author}
  {\bibfnamefont {U.}~\bibnamefont {Schollw\"ock}},\ }\bibfield  {title}
  {\bibinfo {title} {Imaginary-time matrix product state impurity solver for
  dynamical mean-field theory},\ }\href
  {https://doi.org/10.1103/PhysRevX.5.041032} {\bibfield  {journal} {\bibinfo
  {journal} {Phys. Rev. X}\ }\textbf {\bibinfo {volume} {5}},\ \bibinfo {pages}
  {041032} (\bibinfo {year} {2015}{\natexlab{a}})}\BibitemShut {NoStop}%
\bibitem [{\citenamefont {Wolf}\ \emph
  {et~al.}(2015{\natexlab{b}})\citenamefont {Wolf}, \citenamefont {Justiniano},
  \citenamefont {McCulloch},\ and\ \citenamefont
  {Schollw\"ock}}]{WolfChebyPRB2015}%
  \BibitemOpen
  \bibfield  {author} {\bibinfo {author} {\bibfnamefont {F.~A.}\ \bibnamefont
  {Wolf}}, \bibinfo {author} {\bibfnamefont {J.~A.}\ \bibnamefont
  {Justiniano}}, \bibinfo {author} {\bibfnamefont {I.~P.}\ \bibnamefont
  {McCulloch}},\ and\ \bibinfo {author} {\bibfnamefont {U.}~\bibnamefont
  {Schollw\"ock}},\ }\bibfield  {title} {\bibinfo {title} {Spectral functions
  and time evolution from the chebyshev recursion},\ }\href
  {https://doi.org/10.1103/PhysRevB.91.115144} {\bibfield  {journal} {\bibinfo
  {journal} {Phys. Rev. B}\ }\textbf {\bibinfo {volume} {91}},\ \bibinfo
  {pages} {115144} (\bibinfo {year} {2015}{\natexlab{b}})}\BibitemShut
  {NoStop}%
\bibitem [{\citenamefont {Linden}\ \emph {et~al.}(2020)\citenamefont {Linden},
  \citenamefont {Zingl}, \citenamefont {Hubig}, \citenamefont {Parcollet},\
  and\ \citenamefont {Schollw\"ock}}]{Schollwoeck_ImagTimeSolver_2}%
  \BibitemOpen
  \bibfield  {author} {\bibinfo {author} {\bibfnamefont {N.-O.}\ \bibnamefont
  {Linden}}, \bibinfo {author} {\bibfnamefont {M.}~\bibnamefont {Zingl}},
  \bibinfo {author} {\bibfnamefont {C.}~\bibnamefont {Hubig}}, \bibinfo
  {author} {\bibfnamefont {O.}~\bibnamefont {Parcollet}},\ and\ \bibinfo
  {author} {\bibfnamefont {U.}~\bibnamefont {Schollw\"ock}},\ }\bibfield
  {title} {\bibinfo {title} {Imaginary-time matrix product state impurity
  solver in a real material calculation: Spin-orbit coupling in
  $\mathrm{Sr}{}_{2}\mathrm{RuO}{}_{4}$},\ }\href
  {https://doi.org/10.1103/PhysRevB.101.041101} {\bibfield  {journal} {\bibinfo
   {journal} {Phys. Rev. B}\ }\textbf {\bibinfo {volume} {101}},\ \bibinfo
  {pages} {041101} (\bibinfo {year} {2020})}\BibitemShut {NoStop}%
\bibitem [{\citenamefont {Wolf}\ \emph
  {et~al.}(2014{\natexlab{a}})\citenamefont {Wolf}, \citenamefont {McCulloch},
  \citenamefont {Parcollet},\ and\ \citenamefont
  {Schollw\"ock}}]{WolfClusterPhysRevB.90.115124}%
  \BibitemOpen
  \bibfield  {author} {\bibinfo {author} {\bibfnamefont {F.~A.}\ \bibnamefont
  {Wolf}}, \bibinfo {author} {\bibfnamefont {I.~P.}\ \bibnamefont {McCulloch}},
  \bibinfo {author} {\bibfnamefont {O.}~\bibnamefont {Parcollet}},\ and\
  \bibinfo {author} {\bibfnamefont {U.}~\bibnamefont {Schollw\"ock}},\
  }\bibfield  {title} {\bibinfo {title} {Chebyshev matrix product state
  impurity solver for dynamical mean-field theory},\ }\href
  {https://doi.org/10.1103/PhysRevB.90.115124} {\bibfield  {journal} {\bibinfo
  {journal} {Phys. Rev. B}\ }\textbf {\bibinfo {volume} {90}},\ \bibinfo
  {pages} {115124} (\bibinfo {year} {2014}{\natexlab{a}})}\BibitemShut
  {NoStop}%
\bibitem [{\citenamefont {Karp}\ \emph {et~al.}(2020)\citenamefont {Karp},
  \citenamefont {Bramberger}, \citenamefont {Grundner}, \citenamefont
  {Schollw\"ock}, \citenamefont {Millis},\ and\ \citenamefont
  {Zingl}}]{Zingl_ImagTimeSolver}%
  \BibitemOpen
  \bibfield  {author} {\bibinfo {author} {\bibfnamefont {J.}~\bibnamefont
  {Karp}}, \bibinfo {author} {\bibfnamefont {M.}~\bibnamefont {Bramberger}},
  \bibinfo {author} {\bibfnamefont {M.}~\bibnamefont {Grundner}}, \bibinfo
  {author} {\bibfnamefont {U.}~\bibnamefont {Schollw\"ock}}, \bibinfo {author}
  {\bibfnamefont {A.~J.}\ \bibnamefont {Millis}},\ and\ \bibinfo {author}
  {\bibfnamefont {M.}~\bibnamefont {Zingl}},\ }\bibfield  {title} {\bibinfo
  {title} {${\mathrm{sr}}_{2}{\mathrm{moo}}_{4}$ and
  ${\mathrm{sr}}_{2}{\mathrm{ruo}}_{4}$: Disentangling the roles of hund's and
  van hove physics},\ }\href {https://doi.org/10.1103/PhysRevLett.125.166401}
  {\bibfield  {journal} {\bibinfo  {journal} {Phys. Rev. Lett.}\ }\textbf
  {\bibinfo {volume} {125}},\ \bibinfo {pages} {166401} (\bibinfo {year}
  {2020})}\BibitemShut {NoStop}%
\bibitem [{\citenamefont {Bauernfeind}\ \emph {et~al.}(2017)\citenamefont
  {Bauernfeind}, \citenamefont {Zingl}, \citenamefont {Triebl}, \citenamefont
  {Aichhorn},\ and\ \citenamefont {Evertz}}]{FTPS}%
  \BibitemOpen
  \bibfield  {author} {\bibinfo {author} {\bibfnamefont {D.}~\bibnamefont
  {Bauernfeind}}, \bibinfo {author} {\bibfnamefont {M.}~\bibnamefont {Zingl}},
  \bibinfo {author} {\bibfnamefont {R.}~\bibnamefont {Triebl}}, \bibinfo
  {author} {\bibfnamefont {M.}~\bibnamefont {Aichhorn}},\ and\ \bibinfo
  {author} {\bibfnamefont {H.~G.}\ \bibnamefont {Evertz}},\ }\bibfield  {title}
  {\bibinfo {title} {Fork tensor-product states: Efficient multiorbital
  real-time dmft solver},\ }\href {https://doi.org/10.1103/PhysRevX.7.031013}
  {\bibfield  {journal} {\bibinfo  {journal} {Phys. Rev. X}\ }\textbf {\bibinfo
  {volume} {7}},\ \bibinfo {pages} {031013} (\bibinfo {year}
  {2017})}\BibitemShut {NoStop}%
\bibitem [{\citenamefont {Wolf}\ \emph
  {et~al.}(2014{\natexlab{b}})\citenamefont {Wolf}, \citenamefont {McCulloch},\
  and\ \citenamefont {Schollw\"ock}}]{WolfStar}%
  \BibitemOpen
  \bibfield  {author} {\bibinfo {author} {\bibfnamefont {F.~A.}\ \bibnamefont
  {Wolf}}, \bibinfo {author} {\bibfnamefont {I.~P.}\ \bibnamefont
  {McCulloch}},\ and\ \bibinfo {author} {\bibfnamefont {U.}~\bibnamefont
  {Schollw\"ock}},\ }\bibfield  {title} {\bibinfo {title} {Solving
  nonequilibrium dynamical mean-field theory using matrix product states},\
  }\href {https://doi.org/10.1103/PhysRevB.90.235131} {\bibfield  {journal}
  {\bibinfo  {journal} {Phys. Rev. B}\ }\textbf {\bibinfo {volume} {90}},\
  \bibinfo {pages} {235131} (\bibinfo {year} {2014}{\natexlab{b}})}\BibitemShut
  {NoStop}%
\bibitem [{\citenamefont {Haegeman}\ \emph {et~al.}(2016)\citenamefont
  {Haegeman}, \citenamefont {Lubich}, \citenamefont {Oseledets}, \citenamefont
  {Vandereycken},\ and\ \citenamefont {Verstraete}}]{TDVP}%
  \BibitemOpen
  \bibfield  {author} {\bibinfo {author} {\bibfnamefont {J.}~\bibnamefont
  {Haegeman}}, \bibinfo {author} {\bibfnamefont {C.}~\bibnamefont {Lubich}},
  \bibinfo {author} {\bibfnamefont {I.}~\bibnamefont {Oseledets}}, \bibinfo
  {author} {\bibfnamefont {B.}~\bibnamefont {Vandereycken}},\ and\ \bibinfo
  {author} {\bibfnamefont {F.}~\bibnamefont {Verstraete}},\ }\bibfield  {title}
  {\bibinfo {title} {Unifying time evolution and optimization with matrix
  product states},\ }\href {https://doi.org/10.1103/PhysRevB.94.165116}
  {\bibfield  {journal} {\bibinfo  {journal} {Phys. Rev. B}\ }\textbf {\bibinfo
  {volume} {94}},\ \bibinfo {pages} {165116} (\bibinfo {year}
  {2016})}\BibitemShut {NoStop}%
\bibitem [{\citenamefont {Bauernfeind}\ and\ \citenamefont
  {Aichhorn}(2020)}]{TDVP_TTN}%
  \BibitemOpen
  \bibfield  {author} {\bibinfo {author} {\bibfnamefont {D.}~\bibnamefont
  {Bauernfeind}}\ and\ \bibinfo {author} {\bibfnamefont {M.}~\bibnamefont
  {Aichhorn}},\ }\bibfield  {title} {\bibinfo {title} {{Time Dependent
  Variational Principle for Tree Tensor Networks}},\ }\href
  {https://doi.org/10.21468/SciPostPhys.8.2.024} {\bibfield  {journal}
  {\bibinfo  {journal} {SciPost Phys.}\ }\textbf {\bibinfo {volume} {8}},\
  \bibinfo {pages} {24} (\bibinfo {year} {2020})}\BibitemShut {NoStop}%
\bibitem [{\citenamefont {Yang}\ and\ \citenamefont
  {White}(2020)}]{Mingru_BasisExtensionTDVP}%
  \BibitemOpen
  \bibfield  {author} {\bibinfo {author} {\bibfnamefont {M.}~\bibnamefont
  {Yang}}\ and\ \bibinfo {author} {\bibfnamefont {S.~R.}\ \bibnamefont
  {White}},\ }\bibfield  {title} {\bibinfo {title} {Time-dependent variational
  principle with ancillary krylov subspace},\ }\href
  {https://doi.org/10.1103/PhysRevB.102.094315} {\bibfield  {journal} {\bibinfo
   {journal} {Phys. Rev. B}\ }\textbf {\bibinfo {volume} {102}},\ \bibinfo
  {pages} {094315} (\bibinfo {year} {2020})}\BibitemShut {NoStop}%
\bibitem [{\citenamefont {Seth}\ \emph {et~al.}(2016)\citenamefont {Seth},
  \citenamefont {Krivenko}, \citenamefont {Ferrero},\ and\ \citenamefont
  {Parcollet}}]{Seth_2016}%
  \BibitemOpen
  \bibfield  {author} {\bibinfo {author} {\bibfnamefont {P.}~\bibnamefont
  {Seth}}, \bibinfo {author} {\bibfnamefont {I.}~\bibnamefont {Krivenko}},
  \bibinfo {author} {\bibfnamefont {M.}~\bibnamefont {Ferrero}},\ and\ \bibinfo
  {author} {\bibfnamefont {O.}~\bibnamefont {Parcollet}},\ }\bibfield  {title}
  {\bibinfo {title} {Triqs/cthyb: A continuous-time quantum monte carlo
  hybridisation expansion solver for quantum impurity problems},\ }\href
  {https://doi.org/10.1016/j.cpc.2015.10.023} {\bibfield  {journal} {\bibinfo
  {journal} {Computer Physics Communications}\ }\textbf {\bibinfo {volume}
  {200}},\ \bibinfo {pages} {274–284} (\bibinfo {year} {2016})}\BibitemShut
  {NoStop}%
\bibitem [{\citenamefont {Wietek}\ \emph {et~al.}(2020)\citenamefont {Wietek},
  \citenamefont {He}, \citenamefont {White}, \citenamefont {Georges},\ and\
  \citenamefont {Stoudenmire}}]{wietek2020stripes}%
  \BibitemOpen
  \bibfield  {author} {\bibinfo {author} {\bibfnamefont {A.}~\bibnamefont
  {Wietek}}, \bibinfo {author} {\bibfnamefont {Y.-Y.}\ \bibnamefont {He}},
  \bibinfo {author} {\bibfnamefont {S.~R.}\ \bibnamefont {White}}, \bibinfo
  {author} {\bibfnamefont {A.}~\bibnamefont {Georges}},\ and\ \bibinfo {author}
  {\bibfnamefont {E.~M.}\ \bibnamefont {Stoudenmire}},\ }\href@noop {}
  {\bibinfo {title} {Stripes, antiferromagnetism, and the pseudogap in the
  doped hubbard model at finite temperature}} (\bibinfo {year} {2020}),\
  \Eprint {https://arxiv.org/abs/2009.10736} {arXiv:2009.10736
  [cond-mat.str-el]} \BibitemShut {NoStop}%
\bibitem [{\citenamefont {Wietek}\ \emph {et~al.}(2021)\citenamefont {Wietek},
  \citenamefont {Rossi}, \citenamefont {Šimkovic IV~au2}, \citenamefont
  {Klett}, \citenamefont {Hansmann}, \citenamefont {Ferrero}, \citenamefont
  {Stoudenmire}, \citenamefont {Schäfer},\ and\ \citenamefont
  {Georges}}]{wietek2021mott}%
  \BibitemOpen
  \bibfield  {author} {\bibinfo {author} {\bibfnamefont {A.}~\bibnamefont
  {Wietek}}, \bibinfo {author} {\bibfnamefont {R.}~\bibnamefont {Rossi}},
  \bibinfo {author} {\bibfnamefont {F.}~\bibnamefont {Šimkovic IV~au2}},
  \bibinfo {author} {\bibfnamefont {M.}~\bibnamefont {Klett}}, \bibinfo
  {author} {\bibfnamefont {P.}~\bibnamefont {Hansmann}}, \bibinfo {author}
  {\bibfnamefont {M.}~\bibnamefont {Ferrero}}, \bibinfo {author} {\bibfnamefont
  {E.~M.}\ \bibnamefont {Stoudenmire}}, \bibinfo {author} {\bibfnamefont
  {T.}~\bibnamefont {Schäfer}},\ and\ \bibinfo {author} {\bibfnamefont
  {A.}~\bibnamefont {Georges}},\ }\href@noop {} {\bibinfo {title} {Mott
  insulating states with competing orders in the triangular lattice hubbard
  model}} (\bibinfo {year} {2021}),\ \Eprint {https://arxiv.org/abs/2102.12904}
  {arXiv:2102.12904 [cond-mat.str-el]} \BibitemShut {NoStop}%
\bibitem [{\citenamefont {Bertrand}\ \emph {et~al.}(2021)\citenamefont
  {Bertrand}, \citenamefont {Bauernfeind}, \citenamefont {Dumitrescu},
  \citenamefont {Ma\ifmmode~\check{c}\else \v{c}\fi{}ek}, \citenamefont
  {Waintal},\ and\ \citenamefont {Parcollet}}]{Bertrand_QQMC}%
  \BibitemOpen
  \bibfield  {author} {\bibinfo {author} {\bibfnamefont {C.}~\bibnamefont
  {Bertrand}}, \bibinfo {author} {\bibfnamefont {D.}~\bibnamefont
  {Bauernfeind}}, \bibinfo {author} {\bibfnamefont {P.~T.}\ \bibnamefont
  {Dumitrescu}}, \bibinfo {author} {\bibfnamefont {M.}~\bibnamefont
  {Ma\ifmmode~\check{c}\else \v{c}\fi{}ek}}, \bibinfo {author} {\bibfnamefont
  {X.}~\bibnamefont {Waintal}},\ and\ \bibinfo {author} {\bibfnamefont
  {O.}~\bibnamefont {Parcollet}},\ }\bibfield  {title} {\bibinfo {title}
  {Quantum quasi monte carlo algorithm for out-of-equilibrium green functions
  at long times},\ }\href {https://doi.org/10.1103/PhysRevB.103.155104}
  {\bibfield  {journal} {\bibinfo  {journal} {Phys. Rev. B}\ }\textbf {\bibinfo
  {volume} {103}},\ \bibinfo {pages} {155104} (\bibinfo {year}
  {2021})}\BibitemShut {NoStop}%
\end{thebibliography}%
\end{document}